\documentclass[12pt]{article}
\usepackage{draft} 
\usepackage{amsmath,amsfonts, feynmp, epsf}
\usepackage{amssymb}
\usepackage{graphicx}
\usepackage{color}
\usepackage{grffile}
\usepackage{array}
\usepackage{makecell}
\newcolumntype{x}[1]{>{\centering\arraybackslash}p{#1}}
\input epsf

\usepackage{hyperref}
\usepackage{graphicx,color}
\usepackage{cite}
\usepackage{mciteplus}

\begin{document}

\unitlength = .8mm

\begin{titlepage}
\rightline{MIT-CTP-4673}

\begin{center}

\hfill \\
\hfill \\
\vskip 1cm

\title{Supervertices and Non-renormalization Conditions in Maximal Supergravity Theories}

\author{Yifan Wang$^\diamondsuit$ and Xi Yin$^\spadesuit$}

\address{$^\diamondsuit$Center for Theoretical Physics, Massachusetts Institute of Technology, \\
Cambridge, MA 02139 USA
\\
$^\spadesuit$Jefferson Physical Laboratory, Harvard University, \\
Cambridge, MA 02138 USA}

\email{yifanw@mit.edu,
xiyin@fas.harvard.edu}

\end{center}

\abstract{ We construct higher derivative supervertices in an effective theory of maximal supergravity in various dimensions, in the super spinor helicity formalism, and derive non-renormalization conditions on up to 14-derivative order couplings from supersymmetry. These non-renormalization conditions include Laplace type equations on the coefficients of $R^4$, $D^4R^4$, and $D^6R^4$ couplings. We also find additional constraining equations, which are consistent with previously known results in the effective action of toroidally compactified type II string theory, and elucidate many features thereof.
}

\vfill

\end{titlepage}

\eject

\tableofcontents

\section{Introduction}

Over the last few decades, tremendous insights into quantum field theories and quantum gravity were gained through the study of maximally supersymmetric theories \cite{Sen:1994fa,Vafa:1994tf,Hull:1994ys,Witten:1995ex,Maldacena:1997re,Obers:1998fb,Arkani-Hamed:2013jha,Arkani-Hamed:2013kca,Bern:2012di,Bern:2011qn,Bern:2012uf}.
 The constraints of maximal supersymmetry on the dynamics have been investigated extensively \cite{Seiberg:1997ax,Green:1997tv,Berkovits:1997pj,Green:1998by,Pioline:1998mn,Berkovits:1998ex,Paban:1998ea,Paban:1998qy,Sethi:1999qv,Green:2005ba,Basu:2008cf,Chang:2014kma,Chang:2014nwa,Chang:2014jta,Lin:2015zea,Lin:2015ixa,Wang:2015jna, Cordova:2015vwa}, and yet much remains to be understood \cite{Brodel:2009hu,Beisert:2010jx,Bossard:2010bd,Kallosh:2014hga,Bern:1998ug,Bern:2006kd,Bern:2007hh,Bern:2009kd,Bern:2011qn,Korchemsky:2015ssa}. The goal of this paper is to explore the constraints on higher derivative couplings in an effective theory of gravity with 32 supersymmetries, in various spacetime dimensions, from the superamplitude perspective. We will derive a set of non-renormalization conditions on F-term couplings of 8, 12, and 14 derivative orders, extending the result of \cite{Wang:2015jna} on type IIB supergravity in ten dimensions. Our results are in agreement with \cite{Bossard:2014lra,Bossard:2014aea,Bossard:2015uga}, in which the supersymmetry invariants are analyzed using the harmonic superspace formalism. These non-renormalization conditions constrain the quantum effective action of toroidal compactifications of type II superstring theories, and appear to be consistent with, and explain many features of, the previous proposals on these F-term couplings \cite{Green:1997tv, Pioline:1998mn, Green:2005ba}.

The ordinary two-derivative maximal supergravity theory, at the classical level, is by now a well understood subject \cite{Cremmer:1978ds,Brink:1979nt,Brink:1980az,Howe:1981xy}. Higher derivative supergravity theories, on the other hand, are notoriously difficult to handle. In the standard component field Lagrangian formulation, the maximal supersymmetry can only be realized on-shell,\footnote{See however \cite{Cederwall:2010tn} for an elegant off-shell approach based on pure spinor superspace. It is not yet known how to extend this formalism to include $R^4$ terms.} and the supersymmetry transformations of the fields must be deformed to accommodate the higher derivative F-term couplings \cite{Green:1998by,Basu:2008cf,Bossard:2014lra,Bossard:2014aea,Bossard:2015uga}. Furthermore, the Lagrangian description harbors the redundancy of field redefinitions, and often obscures underlying symmetries \cite{Britto:2005fq,Elvang:2013cua,Arkani-Hamed:2013jha,Arkani-Hamed:2013kca}. Important progress, nonetheless, has been made in the on-shell superspace formalism \cite{Howe:1980th,Howe:1981xy,Kallosh:1980fi,Cederwall:2001dx,Bossard:2010bd}.

Both the complication of nonlinear deformations of supersymmetry transformations and the ambiguity of field redefinitions are evaded in the superamplitude approach \cite{Elvang:2010jv,Elvang:2010xn,Elvang:2013cua,Beisert:2010jx}. One key point is that supersymmetry acts linearly on the amplitudes, and the nonlinearity in the supersymmetry transformation of fields are now hidden in the factorization relations among the amplitudes (tree level unitarity). Instead of trying to classify and constrain terms in the Lagrangian, we will focus on constraining local on-shell vertices that obey supersymmetry Ward identities, i.e. supervertices. The supervertices are basic building blocks of superamplitudes, and are in correspondence with possible first order deformations of the theory, by higher dimensional operators that are compatible with supersymmetry. Thinking in terms of supervertices, rather than Lagrangian couplings, allows for a much more efficient way of organizing supersymmetric deformations.

\subsection{On the construction and classification of supervertices}

In order to formulate the supervertices, and superamplitudes in general, we will work in the super spinor helicity formalism in various spacetime dimensions. We adopt the formalism of \cite{CaronHuot:2010rj,Boels:2012ie}, parameterizing 1-particle states of the supergraviton multiplet with the spinor helicity $\zeta$ and Grassmann  variables $\eta$. The basic idea is to split the 32 supercharges into 16 supermomenta $Q\sim \zeta\eta$ and 16 superderivatives $\overline{Q}\sim \zeta \partial/\partial\eta$. This can be done straightforwardly in type IIB supergravity in 10 dimensions, and in all maximal supergravity theories in dimension $D\leq 9$. A typical superamplitude can be written in a way that is manifestly invariant with respect to the $Q$'s, while the nontrivial supersymmetry Ward identities associated with $\overline{Q}$ will be expressed as a set of first order differential equations in $\eta$.

In section 3, we will give the explicit definitions of the super spinor helicity variables $\zeta,\eta$ in spacetime dimensions from 10 to 3, and construct supervertices in each case. One basic class of supervertices, which we refer to as D-term vertices, 
take the form $\delta^{16}(Q) \overline{Q}^{16}{\cal P}(\zeta_i,\eta_i)$, where $\cal P$ stands for a function of the super spinor helicity variables $\zeta_i, \eta_i$ associated with the external particles (labeled by $i=1,\cdots,n$), that is Lorentz invariant and little group invariant. 
The supersymmetry Ward identities are automatically satisfied. In the maximal supergravity theories, the D-term supervertices correspond to Lagrangian deformations at 16-derivative order and higher. We will not have much to say about their coefficients in an effective action, whose moduli dependence is not determined by supersymmetry alone.

The local supervertices that are not of the D-term type will be referred to as F-term supervertices. They arise at 8, 12, and 14-derivative orders,\footnote{We do not know any 16-derivative or higher order supervertices that are not D-terms, but we have not ruled out this possibility.} of the form $\delta^{16}(Q) {\cal F}(\zeta_i,\eta_i)$, where ${\cal F}$ is a polynomial in the super spinor helicity variables. If ${\cal F}(\zeta_i,\eta_i)$ is independent of $\eta$, then it is obviously annihilated by  $\overline{Q}\sim \zeta\partial/\partial\eta$, and therefore the supersymmetry Ward identities are immediately satisfied. Generally, ${\cal F}$ may depend on $\eta_i$ and is annihilated by $\overline{Q}$ up to $Q$-exact terms. 

The F-term supervertices can be constructed in three ways: ${1\over 2}$ BPS vertices of the schematic form ${\cal Q}^{16} {\cal F}$, where ${\cal F}$ is annihilated by half of the 32 supercharges; ${1\over 4}$ BPS vertices of the form ${\cal Q}^{24} {\cal G}$, where ${\cal G}$ is annihilated by 8 supercharges, and ${1\over 8}$ BPS vertices of the form ${\cal Q}^{28}{\cal H}$, where ${\cal H}$ is annihilated by 4 supercharges. The ${1\over 4}$ BPS supervertices only appear at 12-derivative order and higher, and the ${1\over 8}$ BPS supervertices only appear at 14-derivative order and higher.\footnote{One should constrast the terminology of BPS vertices here with the usual notion of BPS operators, as all the BPS vertices refer to fully supersymmetric deformations of the theory. In particular, the ``less BPS" supervertices are special cases of ``more BPS" supervertices.}

In dimension $D\leq 7$, it appears that all ${1\over 2}$ BPS F-term supervertices can be obtained from $\delta^{16}(Q) f(s_{ij})$ by rotating with the compact R-symmetry group $H$. 
In dimensions $D=8$ and 9, there are exceptional ${1\over 2}$ BPS supervertices that do not fall into this class.\footnote{This has to do with the fact that the compact R-symmetry group is semisimple only for $D\leq 7$.} Their explicit constructions will be explained in section 3 and 4.

The construction of the ${1\over 4}$ and ${1\over 8}$ BPS supervertices is more intricate. They do not arise in ten-dimensional type IIB supergravity theories. In section 3, we will construct ${1\over 4}$ and ${1\over 8}$ BPS supervertices at 12 and 14 derivative orders in various dimensions. We find it most convenient to construct these supervertices in $D\leq 5$, due to the very large R-symmetry groups in lower dimensions. In principle, all such supervertices in higher dimensions can be obtained from the lower dimensional cases by the uplifting procedure described in section 4. The existence of these supervertices has important implications on the moduli dependence of $D^4R^4$ and $D^6R^4$ couplings.

\subsection{Moduli dependence and non-renormalization conditions}

As already mentioned, the supervertices classify supersymmetric (higher derivative) deformations of an effective supergravity action. The coefficient of a supervertex is generally a function of the vacuum expectation values of the massless scalars. In maximal supergravity theories, these scalar fields parameterize a coset manifold of the form $G/H$, where $G$ is a noncompact Lie group and $H$ a maximal compact subgroup \cite{Cremmer:1978ds,Cremmer:1978km,deWit:1986mz,Nicolai:1986jk}. Note that while $G$ is a nonlinearly realized symmetry of the {\it two-derivative} supergravity theory, it will be broken explicitly by the higher derivative F-term couplings of consideration.

We would like to constrain the coefficients of these supervertices as a function of the scalar moduli fields. In the context of the effective theory of massless fields of a toroidally compactified type II string theory, this includes determining the coupling dependence of the supervertex, which captures perturbative as well as non-perturbative contributions to the effective action. For this reason, we refer to such supersymmetry constraints as ``non-renormalization conditions".

For instance, the 4-point superamplitude at 8-derivative order cannot factorize through cubic vertices, by momentum power counting. Such an amplitude must be local, and is thus a supervertex of the form $f(\phi^I) \delta^{16}(Q)$, where the coefficient $f(\phi^I)$ is a function of the scalar vevs. In the Lagrangian language, this supervertex is generated by a coupling of the form $f(\phi^I) R^4+\cdots$, where $R^4$ stands for an appropriate contraction of four Riemann tensors, and $\phi^I$ are the moduli fields. By varying the scalar fields, $\phi^I= \phi^I_0 + \delta\phi^I$, we then obtain $(4+k)$-point vertices, which includes scalar-graviton couplings of the schematic form $(\nabla_{\phi_0})^k f(\phi_0) (\delta\phi)^k R^4$. This can also be understood purely from the amplitude perspective, as a relation between an amplitude with soft scalar emissions and the moduli dependence of a lower point amplitude without the external scalars, as explained in section 2.

Importantly, not all such couplings obtained by expanding the scalar dependence of $f(\phi^I)R^4+\cdots$ correspond to local supervertices. We will see that some couplings of the form $(\delta\phi)^k R^4$, or the corresponding on-shell scalar-graviton vertices, do not admit local supersymmetric completions. That is, they are not components of any supervertex, but rather components (or appropriate soft limits) of a nonlocal superamplitude. When this occurs, such a coupling or component vertex is determined (through supersymmetry Ward identities) entirely by the residues of the superamplitude on its poles, which factorize through lower point supervertices.

In the case of the $R^4$ supervertex, this leads to a linear relation between the Hessian of $f(\phi^I)$ and $f(\phi^I)$ itself, which amounts to a set of second order differential equations for $f(\phi^I)$. For the general higher derivative F-terms, such relations among the derivatives of the coefficient functions $f(\phi^I)$ may be nonlinear, depending on the factorization structure of the superamplitude of question.

The factorization structure of the superamplitude determines the differential equation on $f(\phi^I)$ up to the numerical coefficients. While the latter can in principle be fixed by solving supersymmetry Ward identities, in practice this is not easy to do directly, due to the constraints on the spinor helicity variables in general dimensions. In practice, there is a short cut, thanks to superstring perturbation theory. Namely, once the general structure of the differential equation for $f(\phi^I)$ is known, the precise coefficients can be fixed by comparison with {\it any} known set of (sufficiently nontrivial) amplitudes that obey supersymmetry Ward identities and perturbative unitarity.
In most cases, comparison with tree level and possibly one-loop results from type II string theory allows for fixing these differential equations completely.

In section 5, we explicitly analyze these differential equations for the coefficients of the $R^4$, or the corresponding supervertex, in a maximally supersymmetric gravity theory in $6,7,8,9$ dimensions. In addition to the constraints on the Laplacian of the coefficient function, as was proposed in \cite{Obers:1999um,Green:2010wi}, we also find extra constraining relations. Our results agree precisely with the proposals of \cite{Obers:1999um,Green:2010wi} for the effective action of type II string theory compactified on a torus. In these cases, the supersymmetry constraints are sufficiently powerful such that, when combined with the assumption of U-duality, they fix the answer completely.

We also analyze the coefficients of $D^4R^4$ and $D^6R^4$ couplings in $D\leq 5$. It will turn out that, due to the existence of 6-point, ${1\over 4}$ BPS operators, the second order differential equations are not all the constraints, and there will be independent third order differential constraints. The significance of these higher order derivative constraints was recognized in \cite{Bossard:2014lra,Bossard:2014aea,Bossard:2015uga}.\footnote{We thank G. Bossard for emphasizing this to us.}



Our method for constructing supervertices is particularly convenient in lower dimensions, due to the larger R-symmetry group. A useful dimensional uplifting procedure is introduced in section 4. Namely, one may first construct a supervertex in a lower dimensional supergravity theory, which may be viewed as a candidate supervertex of a higher dimensional theory with particle momenta restricted in a sub-spacetime, and then demand Lorentz invariance in the higher dimensional theory.
We use this method to uplift some supervertices to eleven dimensional supergravity, which may be used to constrain the M-theory effective action. Although we do not have a complete proof, it appears that the only F-term supervertices that can be lifted to 11 dimensions are the {\it 4-point} supervertices for $R^4$, $D^4R^4$, and $D^6R^4$ couplings. The coefficients of these terms in the M-theory effective action are previously known by comparison with exact results in compactified theories \cite{Russo:1997mk,Green:1997as,Green:1999pu,Green:2005ba}. This should allow for, in principle, the determination by supersymmetry of up to $R^7$ couplings in the M-theory effective action.\footnote{In practice, this is most easily done by constraining the superamplitudes, order by order in the momentum expansion.}

\section{Soft limits in higher derivative supergravity theories}

A key ingredient in formulating the supersymmetry constraints on an effective action is that we can expand the coupling coefficient in the scalar moduli fields around their vacuum expectation value, and obtain higher point coupling and the corresponding on-shell vertices. The latter will then be constrained by the supersymmetry Ward identity on the amplitudes. For instance, suppose there is an 8-derivative coupling of the form $f(\phi_0^I) R^4$, where $\phi_0^I$ are the vevs of the scalars. By expanding $\phi^I = \phi^I_0 + \delta\phi^I$, we also obtain couplings of the form ${1\over n!}\partial_{I_1}\cdots \partial_{I_n} f(\phi_0) \delta\phi^{I_1}\cdots \delta\phi^{I_n} R^4$. Such a coupling gives rise to a $(4+n)$-point vertex with 4 gravitons and $n$ scalars, which may or may not admit a local supersymmetric completion.

In practice, in order to carry out perturbation theory with the scalar kinetic term governed by a nonlinear sigma model, it is convenient to work in the Riemann normal coordinates centered at the point on the scalar manifold corresponding to the vev $\phi_0^I$. This is so that the cubic terms in the fluctuation fields $\delta\phi^I$ are eliminated. In passing to a general coordinate system $\phi^I$, we can then replace the ordinary derivative at $\phi_0$ by the covariant derivative defined through the Levi-Civita connection on the scalar manifold. Thus, the coefficient of the higher point vertices generated by expanding $f(\phi^I)$ should be given in terms of the covariant derivatives of $f$, namely ${1\over n!}\nabla_{(I_1}\cdots \nabla_{I_n)} f(\phi_0) $.

\begin{figure}[htb]
\centering
\includegraphics[scale=2.1 ]{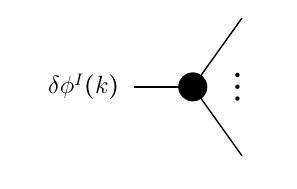}
\raisebox{55pt}{$\xrightarrow{k\rightarrow 0}$}
\includegraphics[scale=2.1]{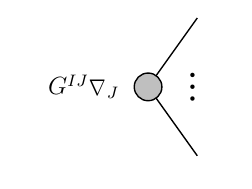}
\caption{Single soft limit of a superamplitude and its relation to the lower point supervertex. }
\label{singlesoft}
\end{figure}

Let $G_{IJ}(\phi)$ be the moduli space metric, and denote by $|\delta\phi^I(k)\rangle$ the 1-particle state created by the field operator $\delta\phi^I$, at momentum $k$. In terms of the amplitudes, the expansion in the fluctuation of scalar moduli fields can be characterized by a soft relation, of the form \cite{Weinberg:1996kr,Weinberg:1966kf} (see Figure \ref{singlesoft})
\ie\label{softr}
\lim_{k\to 0} {\cal A}(\delta\phi^I(k),\{\psi_{i}(p_i)\}) = G^{IJ}(\phi_0)\nabla_{J} {\cal A}(\{\psi_{i}(p_i)\}).
\fe
Here $\psi_{i}(p_i)$ represents other external particles, ${\cal A}$ is the scattering amplitude in the vacuum where $\phi^I$ acquires expectation value $\phi_0^I$. The LHS is the amplitude with the emission of an extra soft scalar $\delta\phi^I$. If $\{\psi_{i}(p_i)\}$ involves the scalar particles created by some $\delta\phi^K$, ${\cal A}(\{\psi_{i}(p_i)\})$ as a function of the scalar vev $\phi_0$ should be viewed as a tensor with respect to coordinate transformations on the scalar manifold, and $\nabla_{J}$ is defined as the appropriate tensor covariant derivative with respect to $\phi_0^J$.

The relation \eqref{softr} can also be understood from the perspective of perturbative string amplitudes. Let us consider type II string tree level amplitude for simplicity. Moving along the moduli space of vacua, the worldsheet CFT is marginally deformed, which to leading order in conformal perturbation theory corresponds to an insertion of $\D\phi^I \int d^2 z\, G_{-{1\over 2}} \overline{G}_{-{1\over 2}}{\cal V}_I$ in the sphere correlation function. Here ${\cal V}_I$ is a superconformal primary of weight $({1\over 2},{1\over 2})$ representing the state $|\D\phi^I(k)\ra$ at momentum $k=0$. This corresponds to the soft emission amplitude of $\D \phi^I$. In the presence of vertex operators of other scalar moduli fields in the correlation function, there are contact terms in the OPEs between them and $G_{-{1\over 2}} \overline G_{-{1\over 2}}{\cal V}_I$, which involves the Levi-Civita connection on the scalar manifold (moduli space of the worldsheet CFT) \cite{Kutasov:1988xb}.  Higher order contributions in conformal perturbation theory involve multiple insertions of the integrated marginal operator, which would lead to more complicated contact terms. 

It is well known that the two-derivative supergravity theory has a nonlinearly realized global symmetry $G$, which is spontaneously broken by the scalar vev to a compact subgroup $H$. The scalar manifold can be identified with the coset space $G/H$.\footnote{The local geometry of the scalar manifold can be understood easily from the scalar 4-point amplitude in the two-derivative supergravity theory. The latter is the scalar component of a 4-point superamplitude of the form $\delta^{16}(Q)/(stu)$ (see section 3 for the super spinor helicity notation). Comparison with the 4-point tree amplitude of the nonlinear sigma model then determines the Riemann tensor of the scalar manifold to be that of a symmetric space.} In this case, the scalars $\phi^I$ may be viewed as Nambu-Goldstone bosons, and the RHS (\ref{softr}) vanishes. This is a version of the single soft pion theorem \cite{Weinberg:1966kf,ArkaniHamed:2008gz}.

The higher derivative couplings considered in this paper, on the other hand, generally break $G$, as well as the subgroup $H$, explicitly. With such couplings viewed as deformations of the Lagrangian, of the schematic form
\ie
{\cal L} = {\cal L}_{(2)} +{\cal O},
\fe 
where ${\cal L}_{(2)}$ is the two-derivative supergravity Lagrangian, the relation (\ref{softr}) can be derived by considering the matrix elements of $\int d^D x\, {\cal O}$ between in and out asymptotic states in the two-derivative supergravity theory. In other words, the amplitude with a soft scalar emission no longer vanishes, but is now related to the derivative of a lower point amplitude with respect to the moduli.

We will also need a soft relation that involves two soft scalar emissions.\footnote{See \cite{Chen:2014cuc,Chen:2014xoa} for recent results of double soft theorems in non-maximal supergravities and the extension to soft fermions.} Generally, the simultaneous double soft limit contains singular terms \cite{Weinberg:1966kf,ArkaniHamed:2008gz}, which involves the structure constants of the group $G$, as well as potential soft graviton poles. However, for our purpose, it suffices to consider the \textit{symmetrized consecutive} soft limit which is non-singular, and takes the form (by applying \eqref{softr} twice)
\ie
{1\over 2} \left( \lim_{k_1\to 0} \lim_{k_2\to 0} + \lim_{k_2\to 0} \lim_{k_1\to 0} \right) {\cal A}(\delta\phi^I(k_1),\delta\phi^J(k_2),\{\psi_{i}(p_i)\}) = G^{(\underline{I}K}(\phi_0) G^{\underline{J})L}(\phi_0) \nabla_{ K}\nabla_{ L}{\cal A}(\{\psi_{i}(p_i)\}).
\fe
In the following sections, we will see that the LHS is often constrained by supersymmetry Ward identities, which then leads to a second order differential equation on the moduli dependence of the amplitude ${\cal A}(\{\psi_{i}(p_i)\})$.

\section{Supervertices in maximal supergravity theories}

We shall begin with the super spinor helicity formalism in ten-dimensional type IIB supergravity \cite{Boels:2012ie}, and review the construction of supervertices in \cite{Wang:2015jna}. The bosonic spinor helicity variable for the supergraviton takes the form $\zeta_{\A A}$, where $\A$ is an $SO(1,9)$ spinor index, and $A=1,\cdots,8$ an $SO(8)$ little group spinor index. $\zeta_{\A A}$ is constrained through the null momentum $p^m$ of the supergraviton via
\ie
p^m\delta_{AB} = \Gamma^m_{\A\B} \zeta_{\A A}\zeta_{\B B},~~~~ \zeta_{\A A}\zeta_{\B A} = {1\over 2} p_m\Gamma^m_{\A\B}.
\fe
One also introduces a set of Grassmann variables $\eta_A$ that transform in the spinor representation of the $SO(8)$ little group. The 1-particle states of the supergraviton multiplet will be identified with polynomials in $\eta_A$. For instance, the dilaton-axion fluctuations $\delta\tau$ and $\delta\overline\tau$ (more precisely, the 1-particle states created by these field operators) are represented by the monomials 1 and $\eta^8\equiv \prod_{A=1}^8 \eta_A$. The 2-form fields are represented by $\eta_A\eta_B$ and ${1\over 6!}\epsilon_{ABA_1\cdots A_6} \eta_{A_1}\cdots \eta_{A_6}$, and the 4-form field and the graviton are represented by degree 4 polynomials in $\eta_A$.

The 32 supercharges that act on the 1-particle state will be split into 16 supermomenta and 16 superderivatives,
\ie
q_\A = \zeta_{\A A}\eta_A,~~~{\rm and}~~\overline{q}_\A = \zeta_{\A A}{\partial\over\partial\eta_A}.
\fe
They obey the supersymmetry algebra 
\ie
\{q_\A, \overline q_\B\} = {1\over 2} p_m \Gamma^m_{\A\B},~~~~
\{q_\A, q_\B\} = \{\overline q_\A, \overline q_\B\}=0. 
\fe
In terms of the supercharges $q^L_\A$ and $q^R_\A$ that arises from the left and right moving sectors of type IIB string theory, we may identify $q_\A = {1\over 2}(q^L_\A + i q^R_\A)$, and $\overline q_\A = {1\over 2}(q^L_\A - i q^R_\A)$. 

An $n$-point superamplitude is the generating function for all amplitudes of $n$ particles in the supergraviton multiplet. It can be expressed as a function of the super spinor helicity variables $\zeta_{i\A A}$, $\eta_{iA}$, where $i=1,\cdots,n$ labels the particles. The 3-point superamplitude is a local supervertex, and is completely fixed by maximal supersymmetry, to be given by the tree level two-derivative supergravity cubic vertex \cite{Boels:2012ie,Wang:2015jna}. For $n\geq 4$, the superamplitude takes the form
\ie\label{agen}
{\cal A} = \delta^{10}(P) \delta^{16}(Q) {\cal F}(\zeta_i,\eta_i).
\fe
Here $P_m$ stands for the total momentum, and $Q_\A = \sum_i q_{i\A}$ is the total supermomenta. 
${\cal F}$ is a function of the super spinor helicity variables. We will be considering mostly tree amplitudes built out of an effective action, that admits a well defined derivative expansion. For this purpose, at a given order in momentum scaling, ${\cal F}$ will be a rational function of the $\zeta_i$'s. 

By construction, (\ref{agen}) is annihilated by the 16 supermomenta $Q_\A$'s, due to the Grassmann  delta function $\delta^{16}(Q)\equiv\prod_\A Q_\A$. The Ward identities associated with the superderivatives $\overline{Q}_\A=\sum_i \overline q_{i\A}$, namely $\overline Q_\A {\cal A}=0$, imposes nontrivial constraints on the function ${\cal F}(\zeta_i,\eta_i)$.

A basic set of building blocks for superamplitudes are on the on-shell supervertices, which are defined to be superamplitudes with no poles in the momenta. The set of linearly independent supervertices are in correspondence with the set of fully supersymmetric infinitesimal deformations of an effective action. An obvious class of supervertices are the D-term vertices of the form 
\ie\label{dter}
\delta^{16}(Q) \overline{Q}{}^{16} {\cal P}(\zeta_i,\eta_i),
\fe
where ${\cal P}$ is an arbitrary polynomial of the super spinor helicity variables that is invariant with respect to Lorentz group, the little groups associated with each of the $n$ external particles, and permutation symmetry on the $n$ particles. The $\delta^{10}(P)$ that enforces momentum conservation will be omitted from now. Note that the $Q_\A$'s and $\overline{Q}_\A$'s are on equal footing in (\ref{dter}).

A supervertex that is not of the D-term type will be referred to as an F-term supervertex. A basic conjecture of \cite{Wang:2015jna}, which can be verified by direct inspection on possible supervertices at low derivative orders, is that the only F-term supervertices in a supergravity theory with type IIB supersymmetry are of the form (for $n\geq 4$)
\ie
\delta^{16}(Q) f(s_{ij}),
\fe
and their CPT conjugates
\ie
f(s_{ij}) \overline{Q}^{16} \prod_{i=1}^n \eta_i^8,
\fe
where $f$ is a polynomial in the Mandelstam variables $s_{ij}=-(p_i+p_j)^2$.

The 4-point supervertex $\delta^{16}(Q)$, for instance, corresponds to the unique supersymmetric completion (at the linearized order) of the 8-derivative coupling of the form $R^4+\cdots$. For $n>4$, the $n$-point 8-derivative supervertex $\delta^{16}(Q)$ corresponds to $(\delta\tau)^{n-4} R^4+\cdots$, and its CPT conjugate corresponds to the coupling $(\delta\overline\tau)^{n-4}R^4+\cdots$. The two-derivative type IIB supergravity theory has a nonlinearly realized $SL(2,\mathbb{R})$ symmetry, whose $U(1)$ subgroup is linearly realized and acts on superamplitudes as
\ie
R = -{1\over 4}\sum_i \left(\eta_{iA}{\partial\over\partial\eta_{iA}}-4 \right).
\fe
This symmetry will be explicitly broken by the supervertices of interest (in particular the ones that arise in string theory), but nonetheless can be used to organize the supervertices and superamplitudes.
In particular, the $n$-point 8-derivative supervertex $\delta^{16}(Q)$ has charge $n-4$ with respect to $R$. 

In the rest of this section, we will consider the reduction of the type IIB super spinor helicity formalism to $D$-dimensions, $3\leq D\leq 9$.\footnote{In less than 3 spacetime dimensions, there is no quantum moduli space of vacua, and amplitudes cannot depend on the scalar field expectation values.} The $D$-dimensional, two-derivative maximal supergravity theory has a nonlinearly realized noncompact symmetry group $G$, out of which a maximal compact subgroup $H$ is linearly realized. We refer to $H$ as the R-symmetry group. The higher derivative supervertices generally transform nontrivially under $H$. We will be able to express the generators of $H$ explicitly as linear operators that act on polynomials in the Grassmann  variables analogous to $\eta_A$. 

A general class of $n$-point 8-derivative F-term supervertices can be constructed by starting with an expression like $\delta^{16}(Q)$ and then rotating by the action of $H$. The ${1\over 2}$ BPS 12-derivative and 14-derivative F-terms can be constructed similarly, from $\delta^{16}(Q)\sum_{i<j} s_{ij}^2$, $\delta^{16}(Q)\sum_{i<j} s_{ij}^3$, and $\delta^{16}(Q)\sum_{i<j<k} s_{ijk}^3$. Note that the latter two expressions give independent $n$-point vertices only when $n\geq 6$. It appears that all ${1\over 2}$ BPS supervertices in $D\leq 7$ dimensions are of this form. In 8 and 9 dimensions, there are some exceptional ${1\over 2}$ BPS supervertices, and a detailed treatment of these supervertices will be given in section 4.

In dimension 9 and below, there are also ${1\over 4}$ BPS supervertices, starting at 12-derivative, 6-point order, of the schematic form $\delta^{16}(Q) \overline{Q}{}^8 {\cal G}(\eta_i)$. In dimension 7 and below, there are ${1\over 8}$ BPS supervertices, of the schematic form $\delta^{16}(Q)\overline{Q}{}^{12} {\cal H}(\eta_i)$. We will give a conjecturally exhaustive construction of the 12-derivative ${1\over 4}$ BPS supervertices in $D\leq 5$, and describe some examples of (but not all) such supervertices in dimensions $6\leq D\leq 9$.

\subsection{9D}

We will denote by $\gamma^m$ the 9 dimensional Gamma matrices, $m=0,\cdots,8$. The spinor helicity variable $\lambda_{\A A}$ of the supergraviton 1-particle state is related to the null momentum $p_m$ by 
\ie\label{ning}
\gamma^m_{\A\B} \lambda_{\A A}\lambda_{\B B} = \delta_{AB} p^m,~~~~ \lambda_{\A A}\lambda_{\A B} = 0.
\fe
Here $\A,\B$ are $SO(1,8)$ spinor indices, and $A,B$ are $SO(7)$ little group spinor indices. Note that (\ref{ning}) is a straightforward reduction of type IIB spinor helicity, with the momentum restricted to nine dimensions. There is also an identity
\ie
\lambda_{\A A} \lambda_{\B A} = {1\over 2} \gamma^m_{\A\B} p_m.
\fe
The supergraviton multiplet will be represented by monomials in the Grassmann  parameters $\eta_A$. There are three little group singlets, which we denote by $1$, $\eta^8\equiv \prod_A \eta_A$, and $\eta^4$ defined using the invariant anti-symmetric 4-form on the spinor representation of $so(7)$. They correspond to the three massless scalars in the 9D supergraviton multiplet. The 9D two-derivative supergravity has a nonlinearly realized $SL(2,\mathbb{R})\times\mathbb{R}^+$ symmetry. Two of the scalars transform under the $SL(2,\mathbb{R})$. Their fluctuations are denoted by $\delta\tau$ and $\delta\overline\tau$, and correspond to the monomials 1 and $\eta^8$ in the super spinor helicity notation. If we view the 9D theory as a reduction of 11D supergravity on a torus, $\delta\tau$ and $\delta\overline\tau$ correspond to deformations of the complex modulus of the compactification torus. The third scalar, which we denote by $\sigma$, parameterizes the area of the compactification torus. Its fluctuation $\delta\sigma$ corresponds to $\eta^4$ in super spinor helicity notation.

The supercharges are represented as
\ie
q_\A = \lambda_{\A A} \eta_A,~~~~
\overline q_\A = \lambda_{\A A} {\partial\over \partial \eta_A}.
\fe
As already mentioned, we can write down an $n$-point, 8-derivative F-term supervertex $\delta^{16}(Q)$ and its CPT conjugate, for all $n\geq 4$. They contain $(\delta\tau)^{n-4}R^4$ and $(\delta\overline\tau)^{n-4}R^4$ couplings. 

We will see in section 4 that there is an exceptional 5-point supervertex that contains $\delta\sigma R^4$. This supervertex will be constructed by showing that a particular supervertex in 8D can be lifted to a Lorentz invariant supervertex in 9D. It appears that this 5-point supervertex together with the $n$-point $(\delta\tau)^{n-4}R^4$ and $(\delta\overline\tau)^{n-4}R^4$ supervertices ($n\geq 4$) described above are the complete set of 8-derivative supervertices in 9D maximal supergravity theories. In section 4, we will see that this classification of 8-derivative supervertices leads to a non-renormalization condition that is precisely consistent with the proposed exact result for $R^4$ coupling in the circle compactification of type II string theory \cite{Green:1997as}.

At 12-derivative order, there is another exceptional 6-point supervertex that contains couplings of schematic form $D^4(\delta\sigma)^2 R^4$. It will be constructed in section 4.2 by uplifting from eight dimensions. Here in nine dimensions, we have not exhausted all the supervertices at 12-derivative order and above. In principle, they should all be attainable from the uplifting procedure, starting from $D\leq 5$.

\subsection{8D}

The 8-dimensional super spinor helicity variables are $\lambda_{AI}$, $\widetilde\lambda_{\dot B}^J$, and Grassmann  variables $\eta^I, \widetilde\eta_I$, where $A$ and $\dot B$ are chiral and anti-chiral spinor indices of $SO(1,7)$, and $I$ is a spinor index of the $SO(6)$ little group (chiral or anti-chiral in the case of lower or upper index). $\lambda$ and $\widetilde\lambda$ are related to the null momentum $p^m$ by\footnote{It follows from the defining relations that $\lambda,\widetilde\lambda$ also obey the Dirac equation $p_m\lambda^I\gamma^m=p_m\gamma^m\widetilde\lambda^J = 0$.}
\ie
p^m \delta_I{}^J = \gamma^m_{{A}{\dot B}} \lambda_{{A}I}\widetilde\lambda_{{\dot B}}{}^J,~~~ \lambda_{{A}I}\widetilde\lambda_{{\dot B}}{}^I = {1\over 2} p_m \gamma^m_{{A}{\dot B}}.
\fe
The 32 supercharges act on the 1-particle states as
\ie
& q_{{A}} = \lambda_{{A} I} \eta^I,~~~ \widetilde q_{{\dot A}} = \widetilde \lambda_{{\dot A}}{}^I \widetilde \eta_I,
\\
& \overline{\widetilde q}_{{A}} = \lambda_{{A} I} {\partial\over \partial\widetilde\eta_I},~~~ \overline q_{{\dot A}} = \widetilde \lambda_{{\dot A}}{}^I {\partial\over \partial\eta^I}.
\fe
Our definition of $\lambda,\widetilde\lambda$ leaves a $GL(4)$ ambiguity, $\lambda_{AI}\to g_I{}^J\lambda_{AJ}$, $\widetilde\lambda_{\dot A}{}^I\to (g^{-1})^I{}_J\widetilde\lambda_{\dot A}{}^J$. An $SU(4)$ subgroup of the $GL(4)$ is identified with the little group in six dimensions. In the Lorentzian signature, the $GL(4)$ is fixed to $SU(4)$ by the 8D reality condition relating $\lambda$ and $\tilde{\lambda}$.\footnote{A similar situation happens for the familiar 4D spinor helicity variables.} In the consideration of the analytic property of S-matrix elements, we need to analytically continue in $\lambda$ and $\widetilde\lambda$, relaxing the reality condition. To do so, we can extend the $GL(4)$ transformation on $\lambda, \widetilde\lambda$ to act on $\eta,\widetilde\eta$ as well, leaving the momentum and supercharges invariant. In constructing supervertices and superamplitudes, we should impose the invariance with respect to the $SL(4)$ subgroup, but not necessarily the diagonal $GL(1)$, which rescales the supermomenta $q$ and $\widetilde q$ oppositely.

The 8-dimensional two-derivative maximal supergravity has a nonlinearly realized global symmetry $G=SL(3)\times SL(2)$, whose linearized realized compact R-symmetry subgroup is $H=SO(3)\times SO(2)$. The generators of $H$ are represented on the 1-particle states by
\ie
& R_+ = \eta\widetilde\eta,~~~ R_- = \partial_\eta \partial_{\widetilde\eta},~~~ 
R_3 = -{1\over 2}( \eta\partial_\eta + \widetilde\eta\partial_{\widetilde\eta})+2,
\\
& R' = {1\over 4}(\eta\partial_\eta - \widetilde\eta\partial_{\widetilde\eta}).
\fe
The 7 scalar particles in the 8D supergraviton multiplet are represented by the monomials $(\eta\widetilde\eta)^m$, $0\leq m\leq 4$, and $\eta^4\equiv {1\over 4!}\epsilon_{IJKL}\eta^I\eta^J\eta^K\eta^L$ and $\widetilde\eta^4$. They transform in the representation ${\bf 5}_0\oplus {\bf 1}_1\oplus {\bf 1}_{-1}$ of $SO(3)\times SO(2)$.

The $n$-point supervertex $\delta^{16}(Q)$ now transforms (as the lowest weight component) in a spin $2(n-4)$ representation of the $SO(3)$, and is uncharged with respect to the $SO(2)$.

There are also a pair of $n$-point supervertices that are charged under the $SO(2)$, that are CPT conjugates of one another, of the form
\ie\label{hspe}
&
V_n^-=\delta^{8}(Q_A) \prod_{A=1}^8 \overline{\widetilde Q}_A \prod_{i=1}^n\widetilde\eta_i^4 
= \delta^{8}(Q_A) \prod_{A=1}^8 \left(\sum_{j=1}^n \lambda_{iAI}{\partial\over\partial \widetilde\eta_{iI}} \right) \prod_{i=1}^n\widetilde\eta_i^4,
\\
&
V_n^+=\delta^{8}(\widetilde Q_{\dot A}) \prod_{A=1}^8 \overline{ Q}_{\dot A} \prod_{i=1}^n \eta_i^4
= \delta^{8}(\widetilde Q_{\dot A}) \prod_{A=1}^8 \left(\sum_{j=1}^n \widetilde\lambda_{i\dot A}{}^I {\partial\over\partial \eta_i{}^I} \right) \prod_{i=1}^n\eta_i^4.
\fe
Indeed, it is easy to verify that $V_n^\pm$ obey supersymmetry Ward identities, namely, they are annihilated by $Q_A, \widetilde Q_{\dot A}, \overline{\widetilde Q}_A$, and $\overline{Q}_{\dot A}$. Note that $V_n^\pm$ are $SO(3)$ invariant, and have charge $\pm  (n-4)$ with respect to the $SO(2)$ generator $R'={1\over 4}\sum_i(\eta_i\partial_{\eta_i} - \widetilde\eta_i\partial_{\widetilde\eta_i})$.

We conjecture that $V_n^\pm$ together with the $SO(3)$ orbit of $\delta^{16}(Q)$ give the complete set of $n$-point 8-derivative supervertices in an 8D maximal supergravity theory. In section 4, we will explain the relation between these vertices and those of the 7D maximal supergravity theories.
We will also see that this classification of supervertices is precisely consistent with the proposal of \cite{Green:2010wi} that in the toroidal compactification of superstring theory to 8 dimensions, the coefficient of $R^4$ coupling in the effective action is the sum of a regularized $SL(3)$ Epstein series and an $SL(2)$ Eisenstein series.

Starting at 12-derivative order, we can write $n$-point ${1\over 4}$ BPS supervertices for $n\geq 6$, of the form
\ie\label{8DquarterBPS}
& \delta^{16}(Q) \overline{Q}{}^8 {\cal P}(\eta_i^4) = \delta^{16}(Q) \prod_{A=1}^8 \left( \sum_{j=1}^n \widetilde\lambda_{i\dot A}{}^I {\partial\over\partial \eta_i{}^I}  \right) {\cal P}(\eta_i^4),~~~{\rm and}
\\
&\delta^{16}(Q) \overline{\widetilde Q}{}^8 {\cal P}(\widetilde\eta_i^4) = \delta^{16}(Q) \prod_{A=1}^8 \left( \sum_{j=1}^n \lambda_{iA I} {\partial\over\partial \widetilde\eta_{iI}}  \right) {\cal P}(\widetilde\eta_i^4).
\fe
Here ${\cal P}(\eta_i^4)$ is a polynomial of the little group invariants $\eta_i^4$, of total degree $4k$ in the $\eta_i$'s, with $2\leq k\leq n-2$.\footnote{The expressions in \eqref{8DquarterBPS} vanish identically for $k$ outside this range.} The case $k=2$ corresponds to the ${1\over 2}$ BPS supervertex of the form $\delta^{16}(Q) \sum_{i<j} s_{ij}^2$. The case $k=n-2$ corresponds to $V_n^\pm \sum_{i<j} s_{ij}^2$, where $V_n^\pm$ are the 8-derivative supervertices given in (\ref{hspe}). We will now focus on the more interesting case of $3\leq k\leq n-3$, which correspond to ${1\over 4}$ BPS couplings.

Let us consider the case $n=6$. We have a pair of 6-point ${1\over 4}$ BPS 12-derivative supervertices, 
\ie\label{qsix}
& \delta^{16}(Q) \overline Q{}^8 \sum_{1\leq i<j<k\leq 6} \eta_i^4 \eta_j^4 \eta_k^4,
\\
& \delta^{16}(Q) \overline{\widetilde Q}{}^8 \sum_{1\leq i<j<k\leq 6} \widetilde\eta_i^4 \widetilde\eta_j^4 \widetilde\eta_k^4.
\fe
It is easy to see that the above supervertices are annihilated by $R_-$, and correspond to the highest weight states in the representations ${\bf 5}_{1}$ and  ${\bf 5}_{-1}$ of $SO(3)\times SO(2)$, respectively. An entire multiplet of 6-point, 12-derivative ${1\over 4}$-BPS supervertices are obtained by applying an $SO(3)$ R-symmetry rotation to (\ref{qsix}). 

Likewise at 14-derivative order, there is a class of $n$-point ${1\over 4}$ BPS supervertices for $n\geq 6$, of the form
\ie\label{ftdd}
& \delta^{16}(Q) \overline{Q}{}^8 \sum_{i<j} s_{ij} {\cal P}_{ij}(\eta_k^4),
\\
&\delta^{16}(Q) \overline{\widetilde Q}{}^8  \sum_{i<j} s_{ij} {\cal P}_{ij}(\widetilde\eta_k^4) .
\fe
where ${\cal P}_{ij}$ are polynomials of the little group invariants $\eta_k^4$, of degree $3\leq k\leq n-3$.
Let us specialize to the case $n=6$, $k=3$, where we can write down a pair of ${1\over 4}$ BPS 14-derivative supervertices,
\ie\label{qqs}
& \delta^{16}(Q) \overline Q{}^8 \sum_{1\leq i<j<k\leq 6} s_{ijk} \eta_i^4 \eta_j^4 \eta_k^4,
\\
& \delta^{16}(Q) \overline{\widetilde Q}{}^8 \sum_{1\leq i<j<k\leq 6} s_{ijk} \widetilde\eta_i^4 \widetilde\eta_j^4 \widetilde\eta_k^4.
\fe
They transform as the highest weight states in the representation ${\bf 5}_{1}$ and  ${\bf 5}_{-1}$ of $SO(3)\times SO(2)$ respectively. Here $s_{ijk} = -(p_i+p_j+p_k)^2=s_{ij}+s_{ik}+s_{jk}$.  A priori, one may write down other polynomials ${\cal P}_{ij}(\eta_k^4)$, but after taking into momentum conservation and permutation invariance, we find that (\ref{qqs}) is the only independent ${1\over 4}$ BPS supervertex of the form (\ref{ftdd}).

\subsection{7D}\label{sec7D}

In 7 dimensions, the spinor helicity variables are denoted $\lambda_{\A I}$, where $\A=1,\cdots,8$ is a spinor index of $SO(1,6)$, and $I=1,\cdots,4$ is a spinor index of the $SO(5)$ little group. $\lambda$ is related to the null momentum by
\ie
\gamma^m_{\A\B}\lambda_{\A I}\lambda_{\B J} = p^m \Omega_{IJ}, ~~~~\lambda_{\A I}\lambda_{\A J}=0,
\fe
and
\ie
\lambda_{\A I}\lambda_{\B J}\Omega^{IJ} = {1\over 2} p_m \gamma^m_{\A\B}.
\fe
Here $\Omega_{IJ}$ is the invariant anti-symmetric 2-form on the spinor representation of $SO(5)$. To describe the supergraviton multiplet, we introduce the Grassmann  variables $\eta^I_a$, where $a=\pm$ is an auxiliary index that may be identified with the spinor index of an $SO(3)$ subgroup of the $SO(5)$ R-symmetry (not to be confused with the little group). The supercharges are represented on the 1-particle states as
\ie
& q_{\A a} = \lambda_{\A I}\eta^I_a,~~~~\overline q_{\A}{}^a = \lambda_{\A I} {\partial\over\partial \eta_{Ia}}.
\fe
The $I,J$ indices are raised and lowered with the invariant tensor $\Omega^{IJ}$ and $\Omega_{IJ}$. There are 14 scalars that parameterize the coset $G/H=SL(5)/SO(5)$. The scalar 1-particle states are represented in the super spinor helicity notation by the little group invariant monomials
\ie
1,~~ \Omega_{IJ} \eta^I_{(a} \eta^J_{b)} ,~~ \Omega_{IJ} \eta^I_{(a} \eta^J_{b)}\Omega_{KL} \eta^K_{(c} \eta^L_{d)},~~\Omega_{IJ} {\partial^2\over \partial\eta_I^{(a} \partial\eta_J^{b)}} \eta^8 , ~~\eta^8.
\fe
They transform in the symmetric traceless 2-tensor representation of $SO(5)_R$.
Acting on a supervertex or superamplitude, the $SO(5)_R$ generators can be expressed explicitly as
\ie\label{sof}
& R^+_{(ab)} =  \sum_i\Omega_{IJ} \eta_i{}^I_{(a} \eta_i{}^J_{b)},~~~R^{-(ab)} = \sum_i \Omega_{IJ} {\partial^2\over \partial\eta_{iI(a} \partial\eta_{iJb)}} ,~~~R^0{}_a{}^b = \sum_i \left( \eta_i{}^I_a {\partial\over\partial \eta_i{}^I_b} - 2\delta_a^b \right).
\fe

The $n$-point F-term vertex $\delta^{16}(Q)$ sits in the lowest weight state of a rank $2(n-4)$ symmetric traceless tensor representation of $SO(5)_R$. The entire $SO(5)_R$ multiplet of $n$-point, 8-derivative supervertices can be produced by acting on $\delta^{16}(Q)$ with $R^+_{(ab)}$. These are the complete set of 8-derivative supervertices in a 7D maximal supergravity theory.

Let us briefly discuss some consequences of this classification of 8-derivative supervertices, which will be elaborated in section 5. The 4-point supervertex $\delta^{16}(Q)$ corresponds to the supersymmetric completion of $R^4+\cdots$ coupling in the effective Lagrangian, as in all other spacetime dimensions.
The 5-point supervertices obtained by acting on $\delta^{16}(Q)$ with $R_{(ab)}^+$ transform in the rank-2 symmetric traceless tensor of $SO(5)_R$. They correspond to couplings of the form $\delta\phi^I R^4+\cdots$, where $\delta\phi^I$ are fluctuations of the 14 scalar moduli fields. The coefficients of these 5-point supervertices, as a function of the moduli $\phi^I$, will be tied to the first order derivatives of the $R^4$ coefficient $f(\phi^I)$ with respect to $\phi^I$.

The 6-point supervertices, obtained by acting on $\delta^{16}(Q)$ with $R_{(ab)}^+$ twice, transform in the rank-4 symmetric traceless tensor of $SO(5)_R$, or $[4,0]$ in the Dynkin label notation. 
These supervertices contain couplings of the form $\delta\phi^I\delta\phi^J R^4$. However, couplings of the form $\delta\phi^I\delta\phi^J R^4$ in the effective Lagrangian give rise to more bosonic vertices with two scalars and four gravitons than the ones that belong to the 6-point supervertex.

A priori, the coupling $\delta\phi^I\delta\phi^J R^4$, or the corresponding bosonic on-shell 6-point vertex, transforms in the representation 
\ie
{\rm Sym}^2[2,0] = [0,0]\oplus [0,4]\oplus [2,0]\oplus [4,0]
\fe 
of $SO(5)_R$. Among the irreducible components, only the $[4,0]$ can be completed to a local 6-point supervertex. The remaining components of the bosonic vertex, in the representation $[0,0]\oplus[0,4]\oplus [2,0]$, must be components of nonlocal superamplitudes, and are determined by the factorization of the latter to lower point supervertices. We will see in section 5.1 that these factorization relations lead to three sets of second order differential equations on the $R^4$ coefficient $f(\phi)$. The $[0,0]$ component is an equation that asserts $f(\phi)$ is an eigenfunction of the Laplacian on the scalar manifold $SL(5)/SO(5)$. This equation has been proposed in \cite{Obers:1999um,Green:2010wi}. The $[0,4]$ and $[2,0]$ components of the equations are additional constraints from supersymmetry.

Starting at 6-point, 12-derivative order, apart from ${1\over 2}$ BPS supervertices in the representation $[4,0]$ with lowest weight state
\ie
\D ^{16}(Q)\sum_{1\leq i<j\leq 6}s_{ij}^2,
\fe
there are ${1\over 4}$ BPS supervertices, that can be constructed similarly to the ones described in the previous subsection. One of them is a straightforward dimensional reduction of (\ref{qsix}),
\ie\label{sev}
\delta^{16}(Q) \,\overline{Q}_-^8 \sum_{1\leq i<j<k<\ell\leq 6} \eta_{i+}^4 \eta_{j+}^4 \eta_{k+}^4 ,
\fe
where 
\ie
& \overline{Q}_-^8\equiv \prod_{\A=1}^8 \overline Q_{\A-} = \prod_{\A=1}^8\left( \sum_{i=1}^6 \lambda_{i\A I} {\partial\over\partial\eta_{I+}}\right),~~~~ \eta_{i+}^4 \equiv {1\over 4!}\epsilon_{IJKL} \eta_{i+}^I \eta_{i+}^J \eta_{i+}^K \eta_{i+}^L.
\fe
(\ref{sev}) is obviously annihilated by $\overline{Q}_{\A+}$ and obeys supersymmetry Ward identities. 

The following combination of the 6-point ${1\over 4}$ BPS supervertex and ${1\over 2}$ BPS supervertex,
\ie
\delta^{16}(Q)\Bigg(\overline{Q}_-^8 \sum_{1\leq i<j<k<\ell\leq 6} \eta_{i+}^4 \eta_{j+}^4 \eta_{k+}^4
-{2^2\times 35\over 12!8! } \sum_{1\leq i\leq j\leq 6}s_{ij}^2(R^+_{(++)})^2\Bigg)
\fe
is annihilated by $R^+_{(++)},R^-_{(+\pm)}, R^0_{(++)}$ and transforms as the lowest weight state in the representation $[0,4]$ of $SO(5)_R$.\footnote{Note that the ${1\over 4}$ BPS supervertex and the ${1\over 2}$ BPS supervertex are separately annihilated by $R^-_{(+\pm)}, R^0_{(++)}$ while only a nontrivial linear combination of them is annihilated by $R^+_{(++)}$.}

The 7D Lorentz group potentially allows for ${1\over 8}$-BPS 14-derivative supervertices of the form $\delta^{16}(Q)\overline{Q}^{12}{\cal G}(\eta_i)$, but they are not straightforward to construct, due to the constrained nature of the spinor helicity variables. It appears simpler to uplift such ${1\over 8}$ BPS supervertices from lower dimensions.

\subsection{6D}

The 6-dimensional spinor helicity variables $\lambda_{A a}$, $\widetilde\lambda^A{}_{\dot a}$ are related to the null momentum $p_{AB}$ (in bispinor notation) via \cite{Cheung:2009dc,Dennen:2009vk,Bern:2010qa}
\ie
p_{AB} = \lambda_{Aa}\lambda_{Bb}\epsilon^{ab},~~~ p^{AB} = {1\over 2}\epsilon^{ABCD}p_{CD} = \widetilde\lambda^A{}_{\dot a}\widetilde\lambda^B{}_{\dot b}\epsilon^{\dot a\dot b}.
\fe
Here $A,B=1,\cdots,4$ are spinor indices of $SO(1,5)$, and $(a,\dot a)$ are $SU(2)\times SU(2)$ little group spinor indices.

To describe the 6D supergraviton multiplet, the 8 Grassmann  variables are organized in the form $\eta^a{}_{a'}, \widetilde\eta^{\dot a}{}_{\dot a'}$, where $a'=1,2$ and $\dot a'=\dot 1,\dot 2$ are auxiliary indices that may be identified with spinor indices of an $SO(3)\times SO(3)$ subgroup of the R-symmetry group $H=SO(5)\times SO(5)$. The 32 supercharges are represented on the 1-particle states as
\ie
& q_{Aa'} = \lambda_{Aa}\eta^a{}_{a'},~~~ \widetilde q^A{}_{\dot a'} = \widetilde\lambda^{A}{}_{\dot a}\widetilde\eta^{\dot a}{}_{\dot a'},
\\
& \overline q_{Aa'} = \lambda_{Aa}{\partial\over\partial \eta_{a}{}^{a'}},~~~ \overline{\widetilde q}{}^A{}_{\dot a'} = \widetilde\lambda^{A}{}_{\dot a}{\partial\over\partial \widetilde\eta_{\dot a}{}^{\dot a'}},
\fe
The R-symmetry group $H$ is generated by
\ie
& (\eta^2)_{a'b'},~~(\partial_\eta^2)_{a'b'},~~\eta_{a'}\partial_{\eta_{b'}} - \delta_{a'}^{b'}, 
\\
& (\widetilde\eta^2)_{\dot a'\dot b'},~~(\partial_{\widetilde\eta}^2)_{\dot a'\dot b'}, ~~\widetilde\eta_{\dot a'}\partial_{\widetilde\eta_{\dot b'}} - \delta_{\dot a'}^{\dot b'}.
\fe
acting on the 1-particle states.

Here we used the notation $(\eta^2)_{a'b'} = \epsilon_{ab} \eta^a{}_{a'} \eta^b{}_{b'}$, $\eta_{a'}\partial_{\eta_b'}=\epsilon_{ab} \eta^a{}_{a'} {\partial\over \partial \eta_{bb'}}$, etc. The $n$-point 8-derivative supervertices are obtained from $\delta^{16}(Q)$ by the $SO(5)\times SO(5)$ rotation, and transform in the representation $[n-4,0;n-4,0]$.

In particular, the 5-point 8-derivative supervertices that contain the scalar-graviton coupling $\delta\phi^I R^4$ transform in the representation $[1,0;1,0]$ of $H=SO(5)\times SO(5)$. This is consistent with the scalars fields parameterizing the coset $SO(5,5)/(SO(5)\times SO(5))$. The 6-point 8-derivative supervertices transform in the representation $[2,0;2,0]$. On the other hand, the coupling $\delta\phi^I \delta\phi^J R^4$ in the effective Lagrangian, or the corresponding bosonic on-shell vertex, transforms with respect to $H$ according to the representation
\ie
{\rm Sym}^2([1,0;1,0]) = [0,0;0,0] \oplus [0,2;0,2] \oplus [0,0;2,0]\oplus [2,0;0,0] \oplus [2,0;2,0].
\fe
Thus we expect the $R^4$ coefficient $f(\phi)$ to be an eigenfunction of the Laplacian on the scalar manifold $SO(5,5)/(SO(5)\times SO(5))$, and further obeys a set of second order differential equations that restricts the $[0,2;0,2]$, $[0,0;2,0]$, and $[2,0;0,0]$ components of the Hessian of $f(\phi)$ (to zero, in fact). These will be examined in detail in section 5.4.

Starting at 12-derivative order, ${1\over 4}$ BPS supervertices can be constructed as before. Here we highlight the construction of a set of ${1\over 8}$ BPS supervertices at 14-derivative order. Consider the $n$-point, 14-derivative supervertex of the form
\ie\label{nds}
\delta^{16}(Q) \overline{Q}{}^8 \overline{\widetilde Q}{}_-^4 {\cal P}(\eta^2_{ia'b'}, \widetilde\eta_{i++}^2 ),
\fe
where 
\ie
& \overline{Q}{}^8 \equiv \prod_{A=1}^4 \overline Q_{A+} \overline Q_{A-},~~~~\overline{\widetilde Q}{}_-^4 \equiv \prod_{A=1}^4 \overline{\widetilde Q}{}^A{}_{-} = \prod_{A=1}^4 \left( \sum_{i=1}^n \widetilde \lambda_i{}^A{}_{\dot a} {\partial\over\partial \widetilde\eta_{i\dot a+}}\right),
\\
& \eta^2_{ia'b'} \equiv \epsilon_{ab} \eta_i{}^a{}_{a'} \eta_i{}^b{}_{b'},~~~~ \widetilde\eta^2_{i++} \equiv \epsilon_{\dot a\dot b} \widetilde\eta_i{}^{\dot a}{}_+ \widetilde\eta_i{}^{\dot b}{}_+.
\fe
${\cal P}$ is a polynomial in the little group invariants, and must be of degree at least 8 in the $\eta$'s and degree at least 4 in the $\widetilde\eta_+$'s. Note that ${\cal P}(\eta^2_{ia'b'}, \widetilde\eta_{i++}^2 )$ is obviously annihilated by $\overline{\widetilde Q}{}^A{}_+$, and consequently (\ref{nds}) obeys the supersymmetry Ward identities. If ${\cal P}$ is of order $\eta^8\widetilde\eta^4$, $ \overline{Q}{}^8 \overline{\widetilde Q}{}_-^4 {\cal P}$ will simply be a Lorentz invariant expression of the momenta that can be expressed in terms of a cubic polynomial in the $s_{ij}$'s. A multiplet of ${1\over 8}$ BPS supervertices can be constructed using ${\cal P}$ of higher degrees in $\eta$ and $\widetilde\eta_+$, and rotating (\ref{nds}) with the R-symmetry group $SO(5)\times SO(5)$.

\subsection{5D}

The 5D spinor helicity variables $\lambda_{Aa}$ are related to the null momentum $p_{AB}$ by
\ie
p_{AB} = \lambda_{Aa}\lambda_{Bb}\epsilon^{ab},~~~ \Omega^{AB}\lambda_{Aa}\lambda_{Bb}\epsilon^{ab}=0.
\fe
Here $A,B=1,\cdots,4$ are spinor indices of $SO(1,4)$, $a,b=1,2$ are spinor indices of the $SO(3)$ little group, and $\Omega^{AB}$ is the invariant anti-symmetric form on the spinor representation of $SO(1,4)$. We then introduce the Grassmann  variables $\eta_{aI}$, where $I=1,\cdots,4$ is an auxiliary index that may be identified with that of an $SU(4)$ subgroup of the R-symmetry group $USp(8)$. The 32 supercharges are represented on the 1-particle states as
\ie
q_{AI} = \lambda_{Aa}\eta^a{}_I,~~~\overline{q}_A{}^I = \lambda_{Aa}{\partial\over\partial \eta_{aI}}.
\fe
The compact R-symmetry group $H=USp(8)$ acts on the 1-particle states as
\ie
R_{IJ}^+ = \eta_{aI} \eta_{bJ}\epsilon^{ab},~~~ R^{-IJ} = {\partial\over\partial \eta_{aI}} {\partial\over\partial \eta_{bJ}} \epsilon_{ab}, ~~~ R^0{}_I{}^J = \eta_{aI} {\partial\over \partial \eta_{aJ}} - \delta_I^J.
\fe
The $n$-point 8-derivative supervertices obtained by rotating $\delta^{16}(Q)$ with the R-symmetry generators $R_{IJ}^+$ transform in the representation $[0,0,0,n-4]$ of $USp(8)$. In particular, the 5-point supervertices obtained by rotating $\delta^{16}(Q)$, which contain the coupling $\delta\phi^{\cal I} R^4$, transform in the 42-dimensional representation $[0,0,0,1]$. The scalar fields $\phi^{\cal I}$ parameterize the 42-dimensional coset manifold $E_{6(6)}/USp(8)$.

While the 6-point 8-derivative supervertices transform in $[0,0,0,2]$ with respect to the $USp(8)$, the couplings $\delta\phi^{\cal I}\delta\phi^{\cal J} R^4$ a priori transform according to
\ie
{\rm Sym}^2[0,0,0,1] = [0,0,0,0]\oplus [0,2,0,0]\oplus [0,0,0,2].
\fe
So in addition to being an eigenfunction of the Laplacian, the coefficient of the $R^4$ coupling $f(\phi)$ obeys a set of second order differential equations that transform in the 308-dimensional representation $[0,2,0,0]$ of $USp(8)$. This extra set of equations asserts that the $[0,2,0,0]$ component of the Hessian $\nabla_{({\cal I}} \nabla_{{\cal J})} f(\phi)$ vanishes.

The 6-point ${1\over 2}$ BPS supervertices in the representation $[0,0,0,2]$ can be analogously constructed at 12 and 14 derivative orders, by applying the $USp(8)$ R-symmetry rotation to $\delta^{16}(Q)\sum s_{ij}^2$, $\delta^{16}(Q)\sum s_{ij}^3$ and $\delta^{16}(Q)\sum s_{ijk}^3$. Interestingly, there are also 6-point, 12-derivative ${1\over 4}$ BPS supervertices that transform in the representation $[0,2,0,0]$. The lowest weight state will be given by a combination of the following ${1\over 4}$ BPS supervertex\footnote{The reason that only $\eta^a{}_{\pm +}$ appears has to do with the reduction of the 8D Grassmann variable $\eta^I$ in the 8D ${1\over 4}$ BPS supervertex \eqref{qsix}. We use $\pm$ indices to label the spins over the internal planes in the decomposition of the spinor representation of the 8D little group $SO(6)$ in lower dimensions. The rightmost subscript $+$ indicates that we are decomposing the chiral spinor $ \eta^I$ of $SO(6)$ as opposed to the anti-chiral spinor $\tilde \eta_I$.\label{footnote:indices}}
\ie
\delta^{16}(Q) \prod_{A=1}^4 \overline{Q}_{A--}\overline{Q}_{A+-}\sum_{1\leq i_1<i_2<i_3\leq 6} \prod_{k=1}^3\left(\eta_{i_k}{}^a{}_{++}\eta_{i_k}{}^b{}_{++}\epsilon_{ab}\right)\left(\eta_{i_k}{}^c{}_{-+}\eta_{i_k}{}^d{}_{-+}\epsilon_{cd}\right).
\fe
with the following ${1\over 2}$ BPS supervertices
\ie
\D^{16}(Q)\sum_{1\leq i< j\leq 6}s_{ij}^2  R^+_{++++} R^+_{-+-+},\quad
\D^{16}(Q)\sum_{1\leq i< j\leq 6}s_{ij}^2  (R^+_{++-+})^2.
\fe
Here we have chosen to write the index $I$ explicitly as $(\pm\pm)$.
The absence of the singlet $[0,0,0,0]$ among the 6-point, 12-derivative supervertices then gives rise to a Laplacian constraint on the moduli dependence of $D^4R^4$.

Likewise, there are 7-point, 12-derivative ${1\over 4}$ BPS supervertices in the representation $[0,2,0,1]$, in addition to the ${1\over 2}$ BPS supervertices in $[0,0,0,3]$. The representations $[0,0,0,1]$, $[0,2,0,0]$, and $[2,0,0,1]$, on the other hand, are absent at this order, which leads to a set of third order differential equations constraining the coefficient of $D^4R^4$.

\subsection{4D}

The 4D spinor helicity variables are the familiar $\lambda_{\A}$, $\widetilde\lambda_\da$, related to the null momentum (in bispinor notation) $p_{\A\db}$ by
\ie
p_{\A\db} = \lambda_{\A}\widetilde\lambda_{\db}.
\fe
The Grassmann  variables are $\eta_I, \widetilde\eta^I$, $I=1,\cdots,4$. $\lambda$ and $\widetilde\eta$ carry charge $+1$ with respect to the $SO(2)$ little group, and $\widetilde\lambda, \eta$ have charge $-1$ with respect to the little group. The supercharges are represented on the 1-particle states as
\ie
& q_{\A I} = \lambda_\A\eta_I,~~~\widetilde q_\da{}^I = \widetilde\lambda_\da \widetilde\eta^I,
\\
& \overline{\widetilde q}_{\A I} = \lambda_\A{\partial\over \partial \widetilde\eta^I},~~~\overline q_{\da}{}^I = \widetilde\lambda_\da{\partial\over \partial \eta_I}.
\fe
The compact R-symmetry group $H=SU(8)$ is generated by the following little group invariants acting on the 1-particle states,
\ie
& R^+{}_I{}^J = \eta_I\widetilde\eta^J, ~~~R^-{}^I{}_J = {\partial\over \partial\eta_I}{\partial\over\partial\widetilde\eta^J},
~~~ M{}_I{}^J =  \eta_I {\partial\over\partial \eta_J} + \widetilde\eta^J {\partial\over\partial \widetilde\eta^I} - \delta_I^J
\\
& N{}_I{}^J = \eta_I {\partial\over\partial \eta_J} - \widetilde\eta^J {\partial\over\partial \widetilde\eta^I} - {1\over 4}\delta_I^J (\eta\partial_\eta - \widetilde\eta\partial_{\widetilde\eta}).
\fe
Note the commutation relation
\ie{}
[R^+{}_I{}^J, R^{-K}{}_L ] = - \delta_I^K\widetilde\eta^J {\partial\over\partial\widetilde\eta^L} + \delta_L^J{\partial\over \partial \eta_K} \eta_I = - {1\over 2} \delta^J_L (M_I{}^K+N_I^K) - {1\over 2} \delta_I^K (M_L{}^J - N_L{}^J).
\fe
The $n$-point 8-derivative supervertices obtained by rotating $\delta^8(Q)$ with $H=SU(8)$ transform in the representation $[0,0,0,n-4,0,0,0]$. In particular, the 5-point supervertices that contain the coupling $\delta\phi^{\cal I} R^4$ transform in the 70-dimensional representation $[0001000]$. The scalars $\phi^{\cal I}$ parameterize the 70-dimensional  coset manifold $E_{7(7)}/SU(8)$.

While the 6-point 8-derivative supervertices transform in $[0002000]$, the couplings $\delta\phi^{\cal I}\delta\phi^{\cal J} R^4$ a priori transform in the representation
\ie
{\rm Sym}^2[0001000] = [0000000]\oplus [0100010]\oplus [0002000].
\fe
Consequently, in addition to being an eigenfunction of the Laplacian, the coefficient $f(\phi)$ of the $R^4$ coupling obeys a set of second order differential equations that transform in the 720-dimensional representation $[0100010]$ of $SU(8)$. This extra set of equations assert that the $[0100010]$ component of the Hessian $\nabla_{({\cal I}}\nabla_{{\cal J})}f(\phi)$ vanishes.

At 12-derivative order, in addition to the 6-point ${1\over 2}$ BPS supervertices in $[0002000]$, there are also ${1\over 4}$ BPS supervertices in $[0100010]$. The latter can be constructed starting from the lowest weight state\footnote{In this case there are no candidate ${1\over 2}$ BPS supervertices with the right charges to mix with the following ${1\over 4}$ BPS supervertex.}
\ie
\delta^{16}(Q) \overline{Q}^8 \sum_{1\leq i<j<k\leq 6} \eta_i^4\eta_j^4\eta_k^4.
\fe
The singlet supervertex is absent.

Likewise, there are 7-point, 12-derivative ${1\over 4}$ BPS supervertices in the representation $[0101010]$, in addition to the ${1\over 2}$ BPS supervertices in $[0003000]$. The representations $[0001000]$, $[0200000]$, $[0000020]$, and $[1001001]$, on the other hand, are absent at this order.

\subsection{3D}

The 3D spinor helicity variables $\lambda_\A$ are related to the null momentum $p_{\A\B}$ by
\ie
p_{\A\B} = \lambda_\A \lambda_\B.
\fe
The Grassmann  variables are now denoted simply as $\eta_A$, where $A=1,\cdots,8$ is an auxiliary index. The little group is a $\mathbb{Z}_2$ under which $\lambda$ and $\eta$ are odd. The supercharges are represented on the 1-particle states as
\ie
q_{\A A} = \lambda_\A \eta_A,~~~ \overline{q}_{\A A} = \lambda_\A {\partial\over\partial \eta_A}.
\fe
The compact R-symmetry group $H=SO(16)$ is generated by the little group invariants
\ie\label{sos}
R^+_{[AB]} = \eta_A \eta_B,~~~ R^-_{[AB]}={\partial\over\partial\eta_A} {\partial\over\partial\eta_B},
~~~ R^0_{AB} = \eta_A {\partial\over\partial \eta_B} - {1\over 2}\delta_{AB}.
\fe
when acting on the 1-particle states.

The $n$-point 8-derivative supervertices obtained from $\delta^8(Q)$ transform in the representation $[0,0,0,0,0,0,0,n-4]$ of $SO(16)$. In particular, the 5-point supervertices that contain $\delta\phi^{\cal I} R^4$ transform in the 128-dimensional spinor representation $[00000001]$. The scalars $\phi^{\cal I}$ parameterize the 128-dimensional coset $E_{8(8)}/SO(16)$.

The 6-point 8-derivative supervertices transform in $[00000002]$. On the other hand, the couplings $\delta\phi^{\cal I}\delta\phi^{\cal J} R^4$ a priori transform according to
\ie
{\rm Sym}^2[00000001] = [00000000]\oplus [00010000]\oplus [00000002].
\fe
So in addition to being an eigenfunction of the Laplacian, the coefficient $f(\phi)$ of the $R^4$ coupling obeys a set of second order differential equations that amounts to the vanishing of the $[00010000]$ component of the Hessian $\nabla_{({\cal I}}\nabla_{{\cal J})} f(\phi)$.

At 12-derivative order, in addition to the 6-point ${1\over 2}$ BPS supervertices in $[00000002]$, there are also ${1\over 4}$ BPS supervertices in $[00010000]$. The lowest weight state state of $[00010000]$ can be constructed by taking a linear combination of the following ${1\over 4}$ BPS supervertex\footnote{See footnote \ref{footnote:indices} for an explanation for the index notation of $\eta_A$.}
\ie
\delta^{16}(Q) \prod_{\A=1,2}\overline Q_{\A+++}\overline Q_{\A+-+}\overline Q_{\A-++}\overline Q_{\A--+} \sum_{1\leq i_1<i_2<i_3\leq 6} \prod_{k=1}^3 \eta_{{i_k}+++}\eta_{{i_k}+-+}\eta_{{i_k}-++}\eta_{{i_k}--+}.
\fe
and the ${1\over 2}$ BPS supervertex
\ie
\D^{16}(Q)\sum_{1\leq i<j\leq 6}s_{ij}^2 R^+_{[+++,+-+]} R^+_{[-++,--+]}.
\fe
Here we have written the index $A$ explicitly as $(\pm\pm\pm)$.
The singlet supervertex is again absent.

Likewise, there are 7-point, 12-derivative ${1\over 4}$ BPS supervertices in the representation $[00010001]$, in addition to the ${1\over 2}$ BPS supervertices in $[00000003]$. The representations $[00000001]$ and $[01000001]$ are absent at this order.

\section{Lifting supervertices}\label{secLift}

\subsection{The general idea}

As seen in the previous section, the supervertices of maximal supergravity theories in $D\leq 7$ dimensions can be conveniently organized according to representations of the compact R-symmetry group. The construction of the general supervertices in 8 and 9 dimensions appear to be more subtle. Furthermore, the super spinor helicity formalism which splits the supercharges into supermomenta and superderivatives cannot be applied directly to type IIA supergravity and the eleven dimensional supergravity. It is nonetheless possible to construct supervertices in these theories, by uplifting supervertices from a lower dimensional theory of maximal supersymmetry.

Supervertices of a higher dimensional supergravity theory can be trivially reduced to (a subset of) supervertices in lower dimensional theories, simply by restricting the momenta of external particles to a lower dimensional sub-spacetime, and identifying an appropriate embedding of the little groups. The reverse procedure is less obvious, since not all supervertices in the lower dimensional theory come from the reduction of supervertices in a higher dimensional supergravity theory. In this section, we introduce a simple method of identifying those supervertices in the lower dimensional theory that can be lifted to a Lorentz invariant supervertex in a higher spacetime dimension. We will then apply this method to explicitly construct supervertices in 8, 9, and 11 dimensions.

Suppose ${\cal A}_n^{(d)}$ is an $n$-point supervertex in $d$ dimensions, and we would like to know whether it lifts to a supervertex in the $D$ dimensional maximal supergravity theory, for some $D>d$. In other words, we want to know whether ${\cal A}_n^{(d)}$ is the dimensional reduction of a $D$-dimensional supervertex. For reasons that will become clear, we shall assume $n\leq d+1$. While this restriction does not allow us to determine whether the most general $n$-point supervertices can be uplifted, for larger values of $n$, it is nonetheless sufficient for the derivation of non-renormalization conditions considered in this paper.

For the generic assignment of the momenta $p_i$ of the $n$ particles in a scattering amplitude, momentum conservation implies that there are $n-1$ independent null momenta. They span an $(n-1)$-plane in the $d$ dimensional spacetime.\footnote{In analyzing tree amplitudes we are free to analytically continue the momenta to complex values, and the spacetime signature will not be essential.} The $d$ dimensional Lorentz invariance implies that the amplitude is invariant with respect to the $SO(d+1-n)$ rotation of the transverse directions to the $(n-1)$-plane, which leaves all momenta $p_i$ fixed. Now, we would like to view this amplitude as that of $n$ supergravitons in $D$ dimensions, subject to the restriction that the momenta $p_i$ lie in a chosen $(n-1)$-plane in $D$ dimensions. If we fix this set of momenta $p_i$, the $D$-dimensional Lorentz invariance then amounts to the invariance of the amplitude with respect to a $SO(D+1-n)$ subgroup. 

\begin{figure}[htb]
\centering
\includegraphics[scale=1.7]{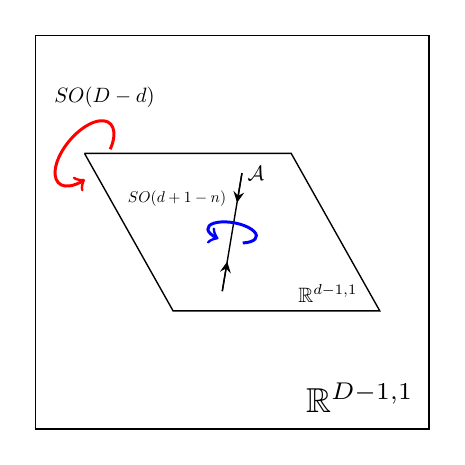}
\caption{Lorentz rotations of an $n$-point amplitude $\cA$ with fixed momenta in a $d$-dimensional sub-spacetime.}
\label{lift}
\end{figure}

The rotation symmetry of the $D-d$ extra dimensions, isomorphic to $SO(D-d)$, embeds into the compact R-symmetry group of the $d$-dimensional supergravity theory. In order for ${\cal A}_n^{(d)}$ to be the restriction of a $D$-dimensional supervertex, we need to at least demand that ${\cal A}_n^{(d)}$ is invariant with respect to the $SO(D-d)$ subgroup of the $d$-dimensional R-symmetry group. With the fixed set of momentum $p_i$, which we assume to be generic, ${\cal A}_n^{(d)}$ is then invariant with respect to an $SO(d+1-n)\times SO(D-d)$ subgroup of the $SO(D+1-n)$ (Figure \ref{lift}). 

The key step is to further demand that $A_n^{(d)}$ is invariant with respect to the remaining $(D-d)(d+1-n)$ generators of the $SO(D+1-n)$. These extra generators, though not manifest, can be realized as differential operators in the Grassmann  variables $\eta_i$ that act on the superamplitude. We denote these generators by $M_a({\bf v})$, where $a=1,\cdots,D-d$, and ${\bf v}$ is a unit vector in $d$ dimensions that obeys ${\bf v}\cdot p_i=0$ for all momenta $p_i$ in the vertex ($i=1,\cdots,n$). They will be constructed explicitly in the next few subsections.


A useful property of $M_a({\bf v})$ is that its commutator with each supercharge is a linear combination of the 32 supercharges, and thus any formal amplitude obtained by acting on ${\cal A}_n^{(d)}$ with $M_a({\bf v})$ will automatically satisfy the supersymmetry Ward identities. The condition for a supervertex ${\cal A}_n^{(d)}$ to lift to a $D$-dimensional Lorentz invariant supervertex is precisely that ${\cal A}_n^{(d)}$ is annihilated by all $M_a({\bf v})$'s.

\subsection{Lifting from 8D to 9D}

Now we consider the problem of lifting supervertices of 8D maximal supergravity to 9D. In this subsection we will denote the 9D super spinor helicity variables by $(\zeta_{\A A}, \eta_A)$, in order to distinguish them from 8D variables.\footnote{In the rest of Section \ref{secLift}, we shall represent the higher dimensional super spinor helicity variables by $(\zeta,\eta)$ and reserve $(\lambda,\theta)$ for the lower dimensional counterparts.} If we restrict the momentum $p$ to 8D, $(\zeta_{\A A}, \eta_A)$ can be decomposed in terms of the 8D super spinor helicity variables $(\lambda_{AI}, \widetilde\lambda_{\dot A}{}^I, \theta^I, \widetilde\theta_I)$ according to
\ie
& \zeta_{\A A} \to (\lambda_{A I}, \lambda_A{}^I=0, \widetilde\lambda_{\dot A I}=0,\widetilde\lambda_{\dot A}{}^I),
\\
& \eta_A \to (\theta^I, \widetilde\theta_I).
\fe
With this restriction, each 9D supervertex reduces to an 8D supervertex. We would like to determine which 8D supervertices arise in this way.

For an $n$-point superamplitude with momenta $p_i$ restricted to 8D, let ${\bf v}$ be a unit vector that obeys ${\bf v}\cdot p_i=0$, for all $i=1,\cdots,n$. We would like to construct the 9D Lorentz generators that rotate the plane spanned by ${\bf v}$ and the extra 9-th dimension. Since the 8D spinor helicity variables $\lambda_i, \widetilde\lambda_i$ obey the Dirac equations
\ie
{\slash\!\!\! p}_{iA\dot B}\widetilde \lambda_{i\dot B}{}^I=0,~~~ {\slash\!\!\! p}_{iA\dot B}\lambda_{iAI}=0,
\fe
where ${\slash\!\!\! p}_{iA\dot B} = p_i^m (\gamma_m)_{A\dot B}$,
the supercharges $q_i, \widetilde q_i, \overline{q}_i, \overline{\widetilde q}_i$ of the $i$-th particle vanish upon contraction with ${\slash\!\!\! p}_i$ as well. For generic null momentum $p_i$, we can then write
\ie
q_i = {\slash\!\!\! p}_i u_i,
\fe
where $u_i$ has the opposite chirality as $q_i$, and is only defined subject to the ambiguity $u_i \sim u_i + {\slash\!\!\! p}_i v_i$.
Similarly, we can define $\widetilde u_i, \overline{u}_i, \overline{\widetilde u}_i$ associated with the other supercharges $\widetilde q_i, \overline{q}_i, \overline{\widetilde q}_i$. Note that the following expression
\ie
q_i {\slash\!\!\! {\bf v}} \overline{\widetilde u}_i = u_i {\slash\!\!\! p}_i {\slash\!\!\! {\bf v}}\overline{\widetilde u}_i
= - u_i {\slash\!\!\! {\bf v}} {\slash\!\!\! p}_i\overline{\widetilde u}_i = - u_i {\slash\!\!\! {\bf v}}\overline{\widetilde q}_i,
\fe
is free of the ambiguity in shifting $u_i$ or $\overline{\widetilde u}_i$. We will define the following linear operators that on the $n$-point amplitude,
\ie
& M^+({\bf v}) = \sum_{i=1}^n q_i {\slash\!\!\! {\bf v}} \overline{\widetilde u}_i  = - \sum_i u_i {\slash\!\!\! {\bf v}}\overline{\widetilde q}_i,
\\
& M^-({\bf v}) = \sum_{i=1}^n \widetilde q_i {\slash\!\!\! {\bf v}} \overline{ u}_i  = - \sum_i \widetilde u_i {\slash\!\!\! {\bf v}}\overline{ q}_i.
\fe
The nonzero commutators of $M^\pm({\bf v})$ with the supercharges are
\ie
& [M^+({\bf v}), \overline Q_{\dot A}] = -{1\over 2} \sum_{i=1}^n ({\slash\!\!\! p}_i {\slash\!\!\! {\bf v}} \overline{\widetilde u}_i )_{\dot A}
= {1\over 2} ({\slash\!\!\! {\bf v}} \overline{\widetilde Q})_{\dot A},
\\
&  [M^+({\bf v}), \widetilde Q_{\dot A}] = - {1\over 2} \sum_{i=1}^n ({\slash\!\!\! p}_i {\slash\!\!\! {\bf v}}  u_i )_{\dot A}
= {1\over 2} ({\slash\!\!\! {\bf v}} Q)_{\dot A},
\\
& [M^-({\bf v}), \overline{\widetilde Q}_A] = -{1\over 2} \sum_{i=1}^n ({\slash\!\!\! p}_i {\slash\!\!\! {\bf v}} \overline{ u}_i )_A
= {1\over 2} ({\slash\!\!\! {\bf v}} \overline{ Q})_A,
\\
&  [M^-({\bf v}), Q_A] = - {1\over 2} \sum_{i=1}^n ({\slash\!\!\! p}_i {\slash\!\!\! {\bf v}}  \widetilde u_i )_A
= {1\over 2} ({\slash\!\!\! {\bf v}} \widetilde Q)_A.
\fe
Thus, if we act on any 8D supervertex with $M^\pm({\bf v})$, the resulting expression still obeys supersymmetry Ward identities with respect to all supercharges.

It will be useful to introduce another operator on the amplitude, 
\ie
M^0 &\equiv 2[M^+({\bf v}), M^-({\bf v})] = -\sum_i \left( u_i {\slash\!\!\! \bf v}{\slash\!\!\! p}_i {\slash\!\!\! \bf v} \overline{u}_i - \widetilde u_i {\slash\!\!\! \bf v}{\slash\!\!\! p}_i {\slash\!\!\! \bf v} \overline{\widetilde u}_i \right)
\\
&=  \sum_i \left( u_i {\slash\!\!\! p}_i \overline{u}_i - \widetilde u_i {\slash\!\!\! p}_i  \overline{\widetilde u}_i \right) 
 =\sum_i \left( q_i \overline{u}_i - \widetilde q_i  \overline{\widetilde u}_i \right)=  \sum_i \left( u_i \overline{q}_i - \widetilde u_i \overline{\widetilde q}_i \right).
\fe
We observe that
\ie
& [M^0, M^\pm({\bf v})] = \pm M^\pm({\bf v}),
\fe
and thus $(M^0,M^+({\bf v}),M^-({\bf v}))$ generate an $su(2)$ algebra.

Now we can identify
\ie
M({\bf v}) = M^+({\bf v}) + M^-({\bf v}) 
\fe
as the generator of rotation in the plane spanned by ${\bf v}$ and the $X^8$ extra dimension direction. If the 8D supervertex is annihilated by $M({\bf v})$ for all ${\bf v}$ perpendicular to the momenta $p_1,\cdots,p_n$, then the vertex is the restriction of a 9D Lorentz invariance vertex. Once again, in making this argument we have assumed $n\leq 9$, so that the set of $n-1$ independent momenta in 9 dimensions can always be rotated into a fixed 8D sub-spacetime.

As an example, let us consider the action of $M({\bf v})$ on the 8-derivative supervertex $\delta^{16}(Q)=\delta^8(Q_A) \delta^8(\widetilde Q_{\dot A})$,
\ie
M({\bf v}) \delta^{16}(Q) &= \left[ - {1\over 2} \sum_i (u_i {\slash\!\!\! {\bf v}} {\slash\!\!\! p}_i)_{\dot A} {\partial\over \partial \widetilde Q_{\dot A}} - {1\over 2} \sum_i (\widetilde u_i {\slash\!\!\! {\bf v}} {\slash\!\!\! p}_i)_{ A} {\partial\over \partial  Q_{ A}} \right] \delta^{16}(Q)
\\
&= \left[  {1\over 2} (Q {\slash\!\!\! {\bf v}} )_{\dot A} {\partial\over \partial \widetilde Q_{\dot A}} + {1\over 2}(\widetilde Q {\slash\!\!\! {\bf v}} )_{ A} {\partial\over \partial  Q_{ A}} \right] \delta^{16}(Q) = 0.
\fe
This shows that $\delta^{16}(Q)$ lifts to a Lorentz invariant supervertex in 9D, as expected.

Now let us consider the 8D 5-point supervertex $V_5^0=\delta^{16}(Q)R_+^2 = \delta^8(Q_A)\delta^8(\widetilde Q_{\dot A}) (\sum_{i=1}^5 \eta_i\widetilde\eta_i)^2$, and $V_5^\pm$ as defined in (\ref{hspe}). Observe that
\ie
& M^+({\bf v}) V_5^0 = 2\delta^{16}(Q) R_+ \sum_{i=1}^5 u_i {\slash\!\!\! {\bf v}} q_i ,
\\
& (M^+({\bf v}))^2 V_5^0 = 2 \delta^{16}(Q)\Big( \sum_{i=1}^5 u_i {\slash\!\!\! {\bf v}} q_i \Big)^2.
\fe
One can verify that $(M^+({\bf v}))^2 V_5^0$ is in fact proportional to $V_5^+$, and similarly $(M^-({\bf v}))^2 V_5^0$ is proportional to $V_5^-$.\footnote{A simple way to see this is to observe that $(M^+({\bf v}))^2 V_5^0$ is of degree 12 in $\eta$ and degree 8 in $\widetilde\eta$, and $V_5^+$ is the unique 5-point vertex of these degrees in $(\eta,\widetilde\eta)$ that obeys the supersymmetry Ward identities.} Thus $(V_5^0,V_5^\pm)$ are three components of a spin-2 multiplet of the $su(2)$ algebra generated by $M^0$ and $M^\pm({\bf v})$. There is a unique linear combination of $(V_5^0,V_5^\pm)$ that is annihilated by $M({\bf v}) = M^+({\bf v}) + M^-({\bf v})$, which lifts to a Lorentz invariant supervertex in 9D. This constructs the 5-point 8-derivative supervertex in 9D that contains the coupling $\delta\sigma R^4$.

On the other hand, the 6-point 8D supervertex of the form $\delta^{16}(Q) R_+^2$ is not annihilated by $M^+({\bf v})$, and its variation under $M^+({\bf v})$ cannot be canceled by the variation of any other 6-point supervertices, as can be seen simply by counting the degrees in $\eta$ and $\widetilde\eta$. It follows that there is no 9D supervertex that contains the coupling $\delta\tau\delta\sigma R^4$.

A similar argument can be used to rule out the 9D supervertices that contain either $\delta\tau\delta\overline\tau R^4$ or $(\delta\sigma)^2 R^4$ couplings. This is because there are only three 6-point supervertices in 8D that are potential candidates for the dimensional reduction of such $U(1)$ neutral supervertices in 9D, namely $\delta^{16}(Q)R_+^4$ and $V_6^\pm$. But unlike the 5-point case, the three supervertices here could only fit into a spin-4 multiplet of the $su(2)$ generated by $M^0, M^\pm({\bf v})$, and no linear combination of the three supervertices can be annihilated by $M({\bf v})=M^+({\bf v})+M^-({\bf v})$. 

In conclusion, inspection of the $M({\bf v})$ transformation on 8D supervertices shows that the only independent $n$-point 9D supervertices at 8-derivative order are $\delta^{16}(Q)$, its CPT conjugate, and the exceptional supervertex in the $n=5$ case, which contains the coupling $\delta\sigma R^4$.

Let us consider another nontrivial example, uplifting a 6-point, 12-derivative supervertex. In 8D, the following 6-point supervertices
\ie
& V_6^+ \sum_{1\leq i<j\leq 6} s_{ij}^2,
\\
& \delta^{16}(Q) \overline{Q}{}^8 R_+^2 \sum_{1\leq i<j<k\leq 6} \eta_i^4 \eta_j^4\eta_k^4 ,
\\
& \delta^{16}(Q) R_+^4 \sum_{1\leq i<j\leq 6} s_{ij}^2,
\\
& \delta^{16}(Q) \overline{\widetilde Q}{}^8 R_+^2 \sum_{1\leq i<j<k\leq 6} \widetilde\eta_i^4 \widetilde\eta_j^4\widetilde\eta_k^4 ,
\\
& V_6^- \sum_{1\leq i<j\leq 6} s_{ij}^2,
\fe
comprise the even $M^0$-eigenvalue components of a spin-4 multiplet of the $su(2)$ algebra generated by $M^0, M^\pm({\bf v})$. A linear combination of them is annihilated by $M({\bf v}) = M^+({\bf v}) + M^-({\bf v})$, and can thus be lifted to a 9D supervertex. This supervertex contains a 6-point coupling of the form $D^4(\delta\sigma)^2 R^4$.

\subsection{The relation between 7D and 8D supervertices}

The connection between our formulation of super spinor helicity variables and supervertices in 7D and 8D requires some explanation. In the usual standard dimensional reduction from 8D to 7D, we take the momenta to lie within a 7-dimensional subspace of the 8D spacetime (transverse to $X^7$ direction), and embed the $SO(5)$ little group of the 7D supergraviton into the $SO(6)$ little group in 8D. Since the chiral and anti-chiral spinors of $SO(1,7)$ reduce to the same spinor representation of $SO(1,6)$, we can write 8D gamma matrix $\gamma^7_{A\dot B}$ as $\delta_{A\dot B}$. The 8D spinor helicity variables $\zeta_{AI}$ and $\zeta_{\dot B}{}^J$ are now subject to the constraint
\ie\label{abi}
\delta_{A\dot B}\zeta_{AI}\widetilde \zeta_{\dot B}{}^J = 0,
\fe
due to the vanishing momentum along $X^7$ direction.
We can then identify them with the 7D spinor helicity variable $\lambda_{AI}$ by 
\ie
\zeta_{A I} = \delta_{A\dot B} \Omega_{IJ}\widetilde\zeta_{\dot B}{}^J = \lambda_{AI},
\fe
and (\ref{abi}) is trivially satisfied.
The Grassmann variable $\eta^I, \widetilde \eta_I$ in the 8D super spinor helicity formalism can be related to the 7D Grassmann  variables $\theta_a^I$($=\Omega^{IJ}\theta_{aI}$), $a=\pm$, through 
\ie
\eta^I = \theta_+^I, ~~~ \widetilde\eta_I =\Omega_{IJ} \theta_-^J.
\fe
The supercharges are then identified as
\ie\label{sid}
& (Q_A, \widetilde Q_{\dot A})\sim (Q_{A+}, \delta_{B\dot A} Q_{B-}),
\\
& (\overline{\widetilde Q}_A, \overline{Q}_{\dot A}) \sim (\overline{Q}_{A-}, \delta_{B\dot A}\overline{Q}_{B+}).
\fe
The $SO(3)\times SO(2)$ compact R-symmetry generators in 8D are identified with a subset of the $SO(5)_R$ generators (\ref{sof}) in 7D, which are
\ie
R^+_{(+-)},~ R^-_{(+-)},~ R^0_{(+-)},~{\rm and}~ R^0_{ab}\epsilon^{ab}.
\fe
The $n$-point supervertex $\delta^{16}(Q)$ in 8D reduces an identical expression in 7D, and hence its $SO(3)$ multiplet maps to the 7D supervertices obtained by acting on $\delta^{16}(Q)$ with the generators $R^+_{(+-)}, R^-_{(+-)}, R^0_{(+-)}$. The $SO(3)$ singlet vertices $V_n^\pm$, which are charged under the $SO(2)$, also reduce to 7D vertices in the $SO(5)$ multiplet of $\delta^{16}(Q)$, but are invariant under $R^+_{(+-)}, R^-_{(+-)}, R^0_{(+-)}$.

There is an alternative route in reducing the 8D spinor spinor helicity formalism to 7D, by identifying
\ie
\eta^I = \theta_+^I, ~~~ \widetilde\eta_I = {\partial\over \partial\theta_-^I}.
\fe
In this formulation, a superamplitude written in the $(\eta^I,\widetilde\eta_I)$-representation can be mapped to a superamplitude in the $\theta_a^I$ representation by a Laplace transform,
\ie
{\cal A}(\eta_i,\widetilde\eta_i) = \int \prod_i d^4\theta_{i-} \,e^{\sum_i \widetilde\eta_{iI}\theta_{i-}^I} {\cal A}(\theta_{i+} = \eta_i,\theta_{i-} ).
\fe
The 8D supercharges would then be identified with the 7D supercharges through
\ie
& (Q_A, \widetilde Q_{\dot A})\sim (Q_{A+}, \delta_{B\dot A} \overline Q_{B-}),
\\
& (\overline{\widetilde Q}_A, \overline{Q}_{\dot A}) \sim (Q_{A-}, \delta_{B\dot A}\overline{Q}_{B+}),
\fe
Compared to (\ref{sid}), this amounts to a different splitting of the 32 supercharges in 7D to supermomenta and superderivatives.
Now the 8D $SO(3)\times SO(2)$ generators reduce to $R^0_{(ab)}$ and $R^0_{ab}\epsilon^{ab}$ of (\ref{sof}), and the 8D supervertices $V_n^\pm$ (\ref{hspe}) reduce to $\delta^{16}(Q)$ and its CPT conjugate in 7D.

\subsection{Lifting to 11D }

The super spinor helicity formalism we have adopted does not permit a straightforward application to superamplitudes in 10 dimensional type IIA supergravity and 11 dimensional supergravity theories, because in these theories there is no Lorentz invariant way of splitting of the 32 supercharges into 16 (mutually anti-commuting) supermomenta and 16 superderivatives. The latter is nonetheless possible if we give up manifest Lorentz invariance. In this subsection, we will construct supervertices in eleven dimensions, by lifting supervertices from 9D and then imposing 11D Lorentz invariance (after having satisfied supersymmetry Ward identities).

The 9D supervertices that are compatible with 11D Lorentz invariance are those annihilated by $M_\pm({\bf v})$, which generate the rotations in the 2-planes spanned by the unit vector ${\bf v}$ (that obeys ${\bf v}\cdot p_i=0$ for all $i$ labeling external particles) and the vectors $\partial_{X^9}\pm i \partial_{X^{10}}$ transverse to the 9D spacetime. In 9D super spinor helicity variables, we have
\ie
& M_+({\bf v}) = \sum_{i=1}^n q_i {\slash\!\!\!\bf v} u_i,
\\
& M_-({\bf v}) =  \sum_{i=1}^n \overline q_i {\slash\!\!\!\bf v} \overline u_i,
\fe
where $u_i$ and $\overline{u}_i$ are defined through
\ie
q_i = {\slash\!\!\! p}_i u_i,~~~\overline q_i = {\slash\!\!\! p}_i \overline u_i.
\fe
Again, $M_\pm({\bf v})$ are well defined, despite that $u_i,\overline{u}_i$ are subject to the ambiguity of shifting by ${\slash\!\!\!p}_i v_i$ or ${\slash\!\!\!p}_i \overline v_i$. The nonzero commutators of $M_\pm({\bf v})$ with the 9D supercharges are
\ie
& [M_+({\bf v}), \overline Q_\A] = {\slash\!\!\! \bf v} Q_\A,~~~ & [M_-({\bf v}), Q_\A] = {\slash\!\!\! \bf v} \overline Q_\A.
\fe
Let us consider the 4-point supervertex $\delta^{16}(Q)$. We have
\ie
M_-({\bf v}) \delta^{16}(Q) = {1\over 4} \sum_i ({\slash\!\!\! p}_i{\slash\!\!\! \bf v})_{\A\B} {\partial\over \partial Q_\A} {\partial\over \partial Q_\B} \delta^{16}(Q)
= {1\over 4} ({\slash\!\!\! P} {\slash\!\!\! \bf v})_{\A\B} {\partial\over \partial Q_\A} {\partial\over \partial Q_\B} \delta^{16}(Q) = 0.
\fe
Similarly, by consideration of CPT conjugation, we see that $\delta^{16}(Q)$ is also annihilated by $M_+({\bf v})$, namely
\ie
M_+({\bf v}) \delta^{16}(Q) = M_+({\bf v}) \overline Q^{16}\prod_{i=1}^4 \eta_i^8 =0.
\fe
Thus we conclude that $\delta^{16}(Q)$ can be lifted to a Lorentz invariant supervertex in 11D. This is nothing but the 11D supervertex that contains the supersymmetric completion of $R^4$ coupling.

In a similar way, all 4-point F-term supervertices of the form $\delta^{16}(Q) {\cal F}(s,t,u)$ in 9D can be lifted to Lorentz invariant supervertices in 11D, which correspond to couplings of the schematic form $D^{2n} R^4+\cdots$.

The $n$-point vertices of the form $\delta^{16}(Q)$, for $n\geq 5$, are still annihilated by $M_-({\bf v})$, but not by $M_+({\bf v})$, neither are they invariant under the $SO(2)$ R-symmetry （(sub)group that rotates the transverse space. Thus, such F-term supervertices in 9D cannot be lifted to Lorentz invariant vertices in 11D. 

The exceptional supervertices in 9D require a more careful analysis. Let us consider the 5-point, 8-derivative supervertex that contains $\delta\sigma R^4$ coupling.
Note that a 9D supervertex that is invariant with respect to $M_+({\bf v}) + M_-({\bf v})$ can be lifted to a supervertex in 10D type IIA supergravity.
From the tree level and 1-loop contribution to $R^4$ effective coupling in type IIA string theory, we know that there is a 5-point 8-derivative supervertex in type IIA supergravity that reduces to a nontrivial linear combination of $\delta\sigma R^4$ and $\delta\tau_2 R^4$ in 9D. To further lift to 11D, we need to demand invariance under $M_+({\bf v})$ and $M_-({\bf v})$ separately. However this is impossible due to the 9D $SO(2)$ R-symmetry charges of these supervertices. Hence we rule out the possibility that a combination of the 9D $\delta\sigma R^4$ supervertex and 5-point $\D^{16}(Q)$ supervertex can be lifted to 11D.


We conjecture that the {\it 4-point} supervertices of the form $\delta^{16}(Q) {\cal F}(s,t,u)$ are in fact the complete set of F-term supervertices in eleven dimensions.\footnote{The argument presented in this section cannot be used to rule out $n$-point F-term supervertices in 11D with $n\geq 11$, which generally cannot be lifted from 9D.}

\section{Non-renormalization conditions from superamplitudes}

The coefficient $f(\phi^I)$ of an F-term supervertex or coupling, as a function of the massless scalar moduli fields $\phi^I$, is generally constrained by supersymmetry. Such constraints take the form of differential equations in $\phi^I$, and are usually referred to as non-renormalization conditions since they can be used to constrain the derivative expansion of a quantum effective action (in the Wilsonian sense). In this section, we wish to establish such non-renormalization conditions on $f(\phi^I)$, in a maximal supergravity theory, along the same line as \cite{Wang:2015jna, Cordova:2015vwa}, in the following steps.

\bigskip

\noindent (1) Expanding $f(\phi^I)$ in the moduli fields $\phi^I$, we obtain higher point {\it component} vertices at the same derivative order. As discussed in section 2, in the amplitude language, this can be understood as a relation between (covariant) derivatives of $f(\phi^I)$ and certain higher point amplitudes with soft scalar emissions.

\bigskip

\noindent (2) These higher point vertices may or may not admit a local supersymmetric completion. That is, they may or may not be a component of a supervertex. If they do not admit a local supersymmetric completion, they must be a component of a superamplitude (that involves soft scalars). If this is the case, the superamplitude of question will be entirely determined by the residues at its poles in the momenta, which are in turn determined by lower point supervertices via factorization, as a consequence of tree level unitarity relation.

\bigskip

\noindent (3) Our classification of F-term supervertices indicates that, typically, only certain special linear combinations of the component vertices obtained by expanding $f(\phi^I)$ to quadratic order in $\delta\phi$ admit local supervertex completions. The linear combinations that cannot be completed as supervertices, which must then be part of superamplitudes that factorize through lower point vertices, will lead to a set of second order differential equation obeyed by $f(\phi^I)$.

\bigskip

\noindent (4) The differential equations obeyed by $f(\phi^I)$ are determined by the general factorization structure of the above mentioned superamplitude on its poles, up to numerical coefficients which can be in principle fixed by supersymmetry Ward identities.

\bigskip

\noindent (5) In practice, rather than directly solving the supersymmetry Ward identities, it suffices to compare the structure of the differential equation for $f(\phi^I)$ with any known set of nontrivial superamplitudes that obey tree level unitarity relations. The superstring perturbation theory provides such a set of amplitudes. Typically, tree level plus possibly one-loop string amplitudes are all that is needed to fix the differential equation obeyed by $f(\phi^I)$. We would like to emphasize that, string perturbation theory is used here as a crutch to nail down the coefficients in the equations. The resulting equations are nonetheless a consequence of supersymmetry alone, and do not depend on the specific string theory.

\bigskip
 
The precise form of these differential equations have been formulated in type IIB supergravity in \cite{Green:1998by,Basu:2008cf,Wang:2015jna}. Below we derive the precise form of the differential equations for the coefficients of various F-term couplings in maximal supergravity theories in lower spacetime dimensions. We begin with the 7D example, where the general features of the supersymmetry constraints on the $R^4$ term are illustrated, and subsequently extend these constraints to 6, 8, and 9 dimensions, where there are some interesting differences due to our classification of 8-derivative supervertices. Note that while in type IIB supergravity, the constraining equation asserts that the coefficient $f(\phi)$ of, say $R^4$ coupling, is an eigenfunction of the Laplacian on the scalar coset manifold $G/H$, in lower spacetime dimensions we will find additional constraints on the Hessian of $f(\phi)$.

We then discuss the consequence of the ${1\over 4}$ BPS supervertices at 12 and 14-derivative orders, focusing on $D\leq 5$ where the R-symmetry representation content is simpler. We will also discuss constraints on independent higher point supervertices at the 14-derivative order, in type IIB supergravity.

\subsection{8-derivative terms in 7D}

Recall that the 14 massless scalars $\phi^I$ of 7D maximal supergravity parameterize the coset $SL(5)/SO(5)$. The possible $n=4$, 5, and 6-point supervertices at 8-derivative order are all generated by applying an $SO(5)_R$ rotation to $\delta^{16}(Q)\equiv\delta^{16}(\sum_{i=1}^n q_i)$. 

\begin{figure}[htb]
\centering

\begin{minipage}{0.3\textwidth}
\centering
\includegraphics[scale=1.2]{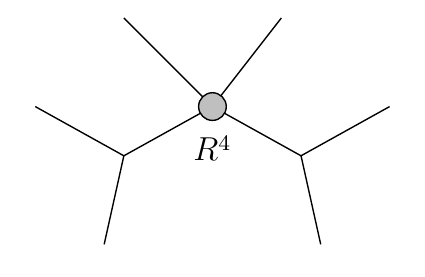}\\
\end{minipage}  
\begin{minipage}{0.3\textwidth}
\centering
\includegraphics[scale=1.2]{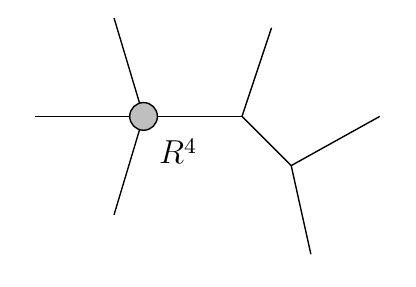}\\
\end{minipage}  
\begin{minipage}{0.3\textwidth}
\centering
\includegraphics[scale=1.2]{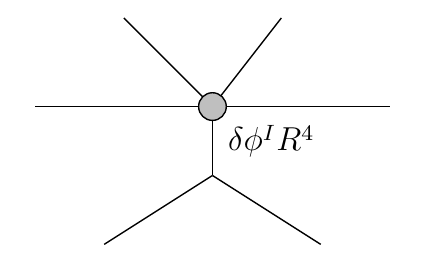}\\
\end{minipage}  
\caption{Factorizations of the two independent 6-point 8-derivative superamplitudes through $R^4$ supervertex and $\D\phi^I R^4$ supervertex respectively.}
\label{7dR4}
\end{figure}

The 4-point supervertex $\delta^{16}(Q)$ contains $R^4$ coupling. We denote the coefficient of this coupling, as a function of the moduli fields, by $f_0(\phi^I)$. Expanding $f_0(\phi^I)$ to second order in the scalar fluctuation $\delta\phi^I$, we find couplings of the form $\delta\phi^I\delta\phi^J R^4$. They transform in the representation
\ie
{\rm Sym}^2[2,0] = [0,0]\oplus [0,4]\oplus[2,0]\oplus [4,0].
\fe
As explained in section 3.3, the only 6-point supervertex at the 8-derivative order transforms in the representation $[4,0]$ of $SO(5)_R$.
The $[0,0]$, $[0,4]$ and $[2,0]$ components of the 6-point scalar-graviton amplitude generated by the coupling $\delta\phi^I\delta\phi^J R^4$ do not admit local supervertex completions, and must be components of a 6-point nonlocal superamplitude, in the corresponding representation of $SO(5)_R$. Such a superamplitude is determined by its factorization through lower point supervertices. By momentum power counting, the factorization must involve precisely one 8-derivative vertex, and a pair of supergravity cubic vertices. Indeed, we have the 4-point $R^4$ supervertex, which is a singlet of $SO(5)_R$, and the 5-point $\delta\phi^I R^4$ supervertex, which transforms in the representation $[2,0]$ (see Figure \ref{7dR4}). There are no 4 or 5-point supervertex that transform in the representation $[0,4]$.

Thus, while the $[4,0]$ component of $\delta\phi^I\delta\phi^J R^4$ is part of a supervertex, whose coefficient is a priori unconstrained by supersymmetry, the $[0,0]$, $[2,0]$, and $[0,4]$ components of $\delta\phi^I\delta\phi^J R^4$ cannot have independent coefficients. It follows from the factorization structure of the 6-point superamplitude at 8-derivative order that there are linear relations between these 6-point couplings and the coefficients of 4 and 5-point supervertices, of the schematic form
\ie
& \big[\nabla_I\nabla_J f_0(\phi)\big]_{[0,0]} \sim f_0(\phi) ,
\\
& \big[\nabla_I\nabla_J f_0(\phi)\big]_{[2,0]} \sim {\partial\over\partial\phi^K}f_0(\phi),
\\
& \big[\nabla_I\nabla_J f_0(\phi)\big]_{[0,4]} =0 .
\fe
The $[0,0]$ component is an equation that asserts $f_0(\phi)$ is an eigenfunction of the Laplacian on the scalar manifold $SL(5)/SO(5)$. The $[2,0]$ and $[0,4]$ components are additional constraints of supersymmetry.

To formulate these equations precisely, let us parameterize the scalar manifold $SL(5)/SO(5)$ by a {\it real  symmetric} $5\times 5$ matrix $g_{ij}$, with $\det g=1$. The $SL(5)$ symmetry of the two-derivative supergravity theory acts by $g\mapsto hgh^{T}$, $h\in SL(5)$. This symmetry is broken explicitly by the higher derivative couplings of interest. We will now write the $R^4$ coefficient $f_0(\phi^I)$ as $f(g)$. 

Since $g_{ij}$ are constrained variables, it will be convenient to formulate the differential equations on $f(g)$ in terms of the variation of $f(g)$ under $g_{ij} \to g_{ij}+\delta g_{ij}$ (as opposed to derivatives with respect to independent variables, which is slightly more cumbersome). Here $\delta g$ is a symmetric real matrix and is subject to the constraint $\det(g+\delta g)=1$, which can be expanded to quadratic order as
\ie\label{ggc}
g^{ij} \delta g_{ij} - {1\over 2} g^{ij} \delta g_{jk} g^{k\ell} \delta g_{\ell i} = {\cal O}((\delta g)^3).
\fe
Now consider the corresponding variation of $f(g)$,
\ie\label{fvar}
\delta f(g) &= \delta g_{ij} f^{ij}(g) + \delta g_{ij} \delta g_{k\ell} f^{ij,k\ell}(g) + {\cal O}((\delta g)^3).
\fe
Due to the constraints (\ref{ggc}), $f^{ij}$ and $f^{ij,k\ell}$ are not unambiguously defined. The ambiguity in $f^{ij}$ can be removed by demanding that $f^{ij}$ is traceless, i.e. $g_{ij} f^{ij}=0$. Likewise, we can fix the ambiguity of $f^{ij,k\ell}$ by demanding that it is traceless with respect to $(ij)$ and $(k\ell)$ respectively.

Now the factorization relation of 6-point amplitudes described above implies the following linear relations\footnote{The $SL(5)$ invariant differential operator we introduced is related to the $SL(5)/SO(5)$ Laplacian in \cite{Green:2010wi} by
\ie
g_{ij}g_{kl}f^{ik,jl}= {1\over 4}\Delta_{SL(5)}f .
\fe }
\ie\label{ggeq}
& g_{ij}g_{k\ell} f^{ik,j\ell}(g) = a f(g),
\\
& g_{k\ell} f^{ik,j\ell}(g) - {1\over 5} g^{ij} g_{k\ell} g_{mn} f^{mk,n\ell}(g) = b   f^{ij}(g) ,
\\
& f^{ij,k\ell}(g) - f^{i\ell, kj}(g) = 0.
\fe
The LHS of these two equations are the projection of $f^{ij,k\ell}$ onto its $[0,0]$, $[2,0]$ and $[0,4]$ components, with respect to $SO(5)_R$.

In the absence of nontrivial $RR$ potential, $g_{ij}$ which captures the $SL(5)/SO(5)$ moduli of type II string theory on $T^3$ has the following $U$-duality invariant parametrization \cite{Green:2010wi}
\ie
g_{ij}=g_7^{-2/5}\begin{pmatrix}
g_7^2 & 0 &  0 \\
 0&   v_3^{-1} & v^{-1}_3 B_{\rm i}^{NS} \\
 0 & v^{-1}_3 B^{NS}_{\rm i}  & v_3^{1\over 3}\tilde g_{\rm ij}-v_3^{-1} B^{NS}_{\rm i}  B^{NS}_{\rm j} 
\end{pmatrix}
\fe
where $v_3=r_1 r_2 r_3/\ell_s^3$ is the volume of $T^3$ and $g_7 =\tau_2^{-1}  v_3^{-1/2}$ is the 7D string coupling. Here $\tilde g_{\rm ij}$ is the $SL(3)$ metric on $T^3$ and $B_{\rm i}^{NS}$ comes from reduction of the two form $B^{NS}$ on 2-cycles of $T^3$ .  


 Comparison with the well known tree level contributions in type II string theory on $T^3$, which is simply $2\zeta(3)g_7^{-12/5}$, then fixes
\ie
a = -{3\over 5},~~~~ b = -{3\over 4}.
\fe
Indeed, the proposal of  \cite{Green:2010wi}, based on a combination of perturbative and instanton computations, and imposing the U-duality symmetry $SL(5,\mathbb{Z})$, asserts that in the toroidal compactification of type II string theory to 7 dimensions, the $R^4$ coefficient as a function of the $SL(5)/SO(5)$ scalars is given by
\ie
f(g) = {\bf E}^{SL(5)}_{[1000];{3\over 2}} (g),
\fe
where ${\bf E}^{SL(5)}_{[1000];s}$ is the $SL(5,\mathbb{Z})$ Epstein series
\ie\label{epstein}
{\bf E}^{SL(5)}_{[1000];s} = \sum_{(m_1,\cdots,m_5)\in\mathbb{Z}^5\backslash \{0\}} {1\over (m^i g_{ij} m^j)^s}.
\fe
Consider the summand in (\ref{epstein}), $f_{(m)}(g) \equiv (m^i g_{ij} m^j)^{-s}$. Under the variation $g\to g+\delta g$, we denote by $f_{(m)}^{ij}$ and $f_{(m)}^{ij,k\ell}$ the first and second order variational coefficients, similarly to (\ref{fvar}). They obey
\ie
f_{(m)}^{ij}&=-{s   \over (m^k g_{kl}m^l)^{s+1}} \Big(m^i m^j-{1\over 5}(m^{k} g_{kl}m^l)g^{ij}\Big),
\\
g_{kl}f_{(m)}^{ik,jl}&={s\over 50(m^k g_{kl}m^l)^{s+1}}\Big(15(s+1)m^i m^j+(s-13)(m^{k} g_{kl}m^l)g^{ij}\Big).
\fe
From this, it is straightforward to verify that the equations (\ref{ggeq}) are obeyed by each summand in the Epstein series (\ref{epstein}), for $s={3\over 2}$.

\subsection{The 6D case}

In this section we extend the analysis of supersymmetry constraints on the $R^4$ coupling in section 5.1 to 6D maximal supergravity theories.

The 6D supergravity scalar coset manifold $SO(5,5)/(SO(5)\times SO(5))$ can be conveniently parameterized by a symmetric $SO(5,5)$ matrix \cite{Maharana:1992my}
\ie
M = \begin{pmatrix} g^{ij} & - g^{ik} b_{kj} \\ b_{ik} g^{kj} & g_{ij} - b_{ik} g^{k\ell} b_{\ell j} \end{pmatrix}.
\fe
Here $g_{ij}$ and $b_{ij}$ are $5\times 5$ real symmetric and anti-symmetric matrices respectively. $g^{ij}$ is the inverse matrix of $g_{ij}$. $M$ obeys 
\ie
M\eta M=\eta,~~~\eta = \begin{pmatrix} 0 & \mathbb{I} \\ \mathbb{I} & 0 \end{pmatrix}.
\fe
An advantage of parameterizing the scalar manifold with $M$ rather than the unconstrained variables $g_{ij}$ and $b_{ij}$ is that $SO(5,5)$ acts linearly on $M$, via\footnote{If we write $\Omega$ in block form, $\Omega = \begin{pmatrix} A & B \\ C & D \end{pmatrix}$, then $X_{ij}\equiv g_{ij}+b_{ij}$ transforms as $X\to (DX +C) (A+BX)^{-1}$.}
\ie
M \to \Omega M \Omega^T.
\fe
Here the $SO(5,5)$ matrix $\Omega$ obeys $\Omega^T \eta\,\Omega = \eta$.

Now consider the expansion around a point $M_0$ in the scalar manifold, and write $M=M_0 + \delta M$. $\delta M_{ij}$ is symmetric and is subject to the constraint
\ie\label{mson}
\delta M \eta M_0 + M_0\eta \delta M = - \delta M \eta \delta M.
\fe
$M_0$ is fixed by a subgroup $H$ of the $SO(5,5)$. In other words, the elements of $H$ are $SO(5,5)$ matrices ${\cal O}$ that obey
\ie\label{omzo}
{\cal O} M_0 {\cal O}^T = M_0.
\fe
By construction, ${\cal O}$ acts on $\delta M$ linearly as well,
\ie
\delta M \to {\cal O}\delta M {\cal O}^T.
\fe
$H$ can be identified with the compact R-symmetry group $SO(5)\times SO(5)$, which acts linearly on the scalar fluctuations.

Let $f(M)$ be the coefficient of the $R^4$ supervertex. We may expand
\ie
f(M) = f(M_0) + \delta M_{ij} f_{(1)}^{ij}(M_0) + \delta M_{ij} \delta M_{k\ell} f_{(2)}^{ij,k\ell}(M_0) + \cdots.
\fe
Due to the constraint on $\delta M$ (\ref{mson}), $f_{(1)}^{ij}$ and $f_{(2)}^{ij,k\ell}$ are subject to the ambiguity of shifting by
\ie
& f_{(1)}^{ij} \to f_{(1)}^{ij} + (\eta M_0 N)^{ji} + (NM_0\eta)^{ji},
\\
& f_{(2)}^{ij,k\ell} \to f_{(2)}^{ij,k\ell} + {1\over 4} (\eta^{ik} N^{j\ell}+ \eta^{jk} N^{i\ell}+\eta^{i\ell} N^{jk}+ \eta^{j\ell} N^{ik}) 
\\
&~~~~~~ + \left[ (\eta M_0 P)^{ji} + (P M_0\eta)^{ji} \right] Q^{k\ell}+ \left[ (\eta M_0 P)^{\ell k} + (P M_0\eta)^{\ell k} \right] Q^{ij},
\fe
where $N$, $P$, $Q$ are arbitrary symmetric matrices. Below we will fix this shift ambiguity and then formulate the differential constraints on $f(M)$ as algebraic relations on the variational coefficients $f^{ij,k\ell}_{(2)}$.

Since $G=SO(5,5)$ is a global symmetry of the two-derivative supergravity theory in 6D, the differential constraining equations on the $R^4$ coupling coefficient $f(M)$ is covariant with respect to the action of $G$, which acts on the scalar manifold as a transitive isometry. Thus, it suffices to examine the constraints on the Hessian of $f(M)$ at a single point on the scalar manifold. For convenience we choose to work with the point $M_0 = \mathbb{I}_{10\times 10}$. Define the $10\times 10$ orthogonal matrix
\ie
S = {1\over \sqrt{2}} \begin{pmatrix} \mathbb{I} & -\mathbb{I} \\ \mathbb{I} & \mathbb{I} \end{pmatrix},
\fe
so that
\ie
S^T \eta S = \begin{pmatrix} \mathbb{I} & 0 \\ 0 & -\mathbb{I} \end{pmatrix} \equiv\widetilde\eta.
\fe
Also define $\widetilde{\delta M} \equiv S^T\delta M S$, so that the constraint (\ref{mson}) can be written as
\ie\label{mcon}
\widetilde{\delta M} \widetilde\eta + \widetilde\eta \widetilde{\delta M} = - \widetilde{\delta M}\widetilde\eta \widetilde{\delta M}.
\fe
In terms of an expansion in $\widetilde{\delta M}$, the first and second order variational coefficients of $f(M)$, denoted by $\widetilde f_{(1)}^{ij}$, $\widetilde f_{(2)}^{ij,k\ell}$, are defined by
\ie
f(M) = f(M_0) + \widetilde{\delta M}_{ij}\widetilde f_{(1)}^{ij}(M_0) + \widetilde{\delta M}_{ij} \widetilde{\delta M}_{k\ell} \widetilde f_{(2)}^{ij,k\ell}(M_0) + \cdots.
\fe
A general element of $H\simeq SO(5)\times SO(5)$, which leaves $M_0=\mathbb{I}_{10\times 10}$ invariant under (\ref{omzo}), takes the form
\ie
{\cal O} = S\begin{pmatrix} A & 0 \\ 0 & B \end{pmatrix} S^T = {1\over 2} \begin{pmatrix} A+B & A-B\\ A-B & A+B \end{pmatrix},~~~A,B\in SO(5).
\fe
The $SO(5,5)$ vector index $i$ on $\widetilde f^{ij}_{(1)}$ and $\widetilde f^{ij,k\ell}_{(2)}$ can then be decomposed into a pair of $SO(5)\times SO(5)$ vector indices, which we denote by $a$ and $\dot a$. 

We now fix the ambiguity in $\widetilde f_{(1)}$ by demanding that the only non-vanishing components of $\widetilde f_{(1)}$ are $(ij)=(a\dot a)$ and $(\dot aa)$, i.e. $\widetilde f_{(1)}$ is block-off-diagonal. Similarly, we fix the ambiguity in $\widetilde f_{(2)}$ by demanding that the only non-vanishing components of $\widetilde f_{(2)}^{ij,k\ell}$ are $(ij),(k\ell) = (a\dot a)$ and $(\dot aa)$. The independent components are $\widetilde f_{(1)}^{a\dot a}$ and $\widetilde f_{(2)}^{a\dot a,b\dot b}$. The former transforms in $[1,0;1,0]$ of $SO(5)\times SO(5)$, whereas the latter a priori transforms in ${\rm Sym}^2[1,0;1,0]=[0,0;0,0]\oplus [0,2;0,2]\oplus[2,0;0,0]\oplus[0,0;2,0]\oplus [2,0;2,0]$. As discussed in section 2.4, the differential constraints amounts to the vanishing of the $[0,2;0,2]$, $[2,0;0,0]$, and $[0,0;2,0]$ components of  $\widetilde f_{(2)}^{a\dot a,b\dot b}$, and that the singlet component of $\widetilde f_{(2)}^{a\dot a,b\dot b}$ is proportional to $f(M_0)$. Explicitly, these conditions can be written as (at $M_0=\mathbb{I}$)
\ie\label{cdiff}
& \delta_{ab} \delta_{\dot a\dot b} \widetilde f_{(2)}^{a\dot a,b\dot b}(\mathbb{I}) = c  f(\mathbb{I}),
\\
& \delta_{ab} \widetilde f^{a\dot a,b\dot b} (\mathbb{I}) = \delta_{\dot a\dot b} \widetilde f^{a\dot a,b\dot b} (\mathbb{I}) = 0,
\\
& \widetilde f_{(2)}^{a\dot a,b\dot b}(\mathbb{I}) = f_{(2)}^{a\dot b, b\dot a}(\mathbb{I}) .
\fe
The proportionality constant $c$ is in principle fixed by supersymmetry Ward identities. As before, we can determine it simply by comparison with tree level string amplitudes.

The constraint (\ref{mcon}) implies that to linear order in $\delta M$, $\widetilde{\delta M}$ is block-off-diagonal. To quadratic order in $\delta M$, we have
\ie
\widetilde{\delta M}_{ab} = {1\over 2} \widetilde{\delta M}_{a\dot c} \widetilde{\delta M}_{b\dot c},~~~~ \widetilde{\delta M}_{\dot a\dot b} = {1\over 2}\widetilde{\delta M}_{c\dot a} \widetilde{\delta M}_{c\dot b}.
\fe
Let $v$ be an arbitrary $SO(5,5)$ vector, and $u\equiv S^T v$. We can write
\ie
v^T \delta M v &= u^T \widetilde{\delta M} u = \widetilde{\delta M}_{ab} u^a u^b +\widetilde{\delta M}_{\dot a\dot b} u^{\dot a} u^{\dot b} + 2\widetilde{\delta M}_{a\dot b} u^a u^{\dot b}
\\
&= 2\widetilde{\delta M}_{a\dot b} u^a u^{\dot b} + {1\over 2} \widetilde{\delta M}_{a\dot c} \widetilde{\delta M}_{b\dot c} u^a u^b + {1\over 2} \widetilde{\delta M}_{c\dot a} \widetilde{\delta M}_{c\dot b} u^{\dot a} u^{\dot b} + {\cal O}((\delta M)^3).
\fe
Now expanding the function
\ie\label{EpsteinFs}
F_s(M) \equiv (v^T M v)^{-s}
\fe
with $M=\mathbb{I}+\delta M$,
\ie
F_s(\mathbb{I}+\delta M) &= (u^T u)^{-s} - 2 s (u^T u)^{-s-1} \widetilde{\delta M}_{a\dot b} u^a u^{\dot b} +  
2s(s+1) (u^Tu)^{-s-2}\widetilde{\delta M}_{a\dot b} \widetilde{\delta M}_{c\dot d} u^a u^{\dot b}u^c u^{\dot d}
\\
&~~~ - {s\over 2} (u^T u)^{-s-1}\left( \widetilde{\delta M}_{a\dot c} \widetilde{\delta M}_{b\dot c} u^a u^b +\widetilde{\delta M}_{c\dot a} \widetilde{\delta M}_{c\dot b} u^{\dot a} u^{\dot b} \right) + {\cal O}((\delta M)^3).
\fe
In particular, we can extract the coefficient of $\widetilde{\delta M}_{a\dot b} \widetilde{\delta M}_{c\dot d}$,
\ie
\widetilde F_{s\,(2)}^{a\dot b,c\dot d} =  F_s(\mathbb{I}) \left[ {2s(s+1)\over (u^T u)^2} u^a u^{\dot b} u^c u^{\dot d} - {s\over 2u^T u} \left( u^a u^c \delta^{\dot b\dot d} + u^{\dot b} u^{\dot d} \delta^{ac} \right) \right].
\fe
Note that $\widetilde F_{s\,(2)}^{a\dot b,c\dot d}$ obeys
\ie\label{taa}
& \delta_{ac} \widetilde F_{s\,(2)}^{a\dot b,c\dot d} - {1\over 5} \delta_{ac}\delta^{\dot b\dot d} \delta_{\dot e\dot f}\widetilde F_{s\,(2)}^{a\dot e,c\dot f}  
\\
&~~~= \left[ {s(2s-3) \delta_{ac} u^a u^c\over (u^T u)^2} + {5s (\delta_{ac}u^au^c - \delta_{\dot a\dot c} u^{\dot a} u^{\dot c}) \over 2(u^T u)^2} \right] \left( u^{\dot b} u^{\dot d} - {1\over 5} \delta^{\dot b\dot d} \delta_{\dot e\dot f} u^{\dot e} u^{\dot f} \right)  F_s(\mathbb{I}),
\fe
as well as a similar equation with the dotted and undotted $SO(5)$ indices exchanged. In order for $F_s(M)$ to obey the second equation of (\ref{cdiff}), which amounts to the vanishing of (\ref{taa}), we need $s={3\over 2}$, and 
\ie\label{tbb}
&\delta_{ac} u^a u^c - \delta_{\dot a\dot c} u^{\dot a}u^{\dot c} = u^T \widetilde\eta u = 0
~~ \Leftrightarrow ~~v^T \eta v = 0.
\fe
When this condition is satisfied, we have
\ie\label{6DR4}
& \delta_{ac}\delta_{\dot b\dot d}\widetilde F_{{3\over 2}\,(2)}^{a\dot b,c\dot d} =-{15\over 8} F_{3\over 2}(\mathbb{I}), 
\\
& \widetilde F_{s\,(2)}^{a\dot b,c\dot d} = F_{s\,(2)}^{a\dot d,c\dot b} .
\fe
Indeed, the string tree level amplitude takes the form of \eqref{EpsteinFs} with $s=3/2$, and a particular $SO(5,5)$ vector $v$ that obeys (\ref{tbb}) \cite{Green:2010wi}, which satisfies all the equations in \eqref{cdiff}, and fixes the constant $c=-15/8$.

The proposal of \cite{Obers:1999um} is that $f(M)$ is an $SO(5,5;\mathbb{Z})$ Eisenstein series that is a sum of terms of the form $F_{3\over 2}(M)$, over charge lattice vectors $v$ that are subject to the constraint $v^T \eta v=0$.\footnote{The constraint $v^T\eta v=0$ is the equivalent to the restriction (3.5b) of \cite{Obers:1999um} in the summation that defines the $SO(5,5;\mathbb{Z})$ Eisenstein series.} As  seen above, this precisely agrees with (\ref{cdiff}), with $c=-15/8$. 

\subsection{The 8D case}

As already seen, the classification of F-term supervertices in 8D and 9D is slightly more intricate than in lower dimensions. In particular, not all F-term supervertices at a given derivative order fall into a single orbit of the compact R-symmetry group. This leads to some interesting features in the supersymmetry non-renormalization conditions. We illustrate this in the example of 4-point 8-derivative coupling, $f(\phi)\delta^{16}(Q)$.

As already explained in section 3.2, in 8D maximal supergravity, the $n$-point vertices at 8-derivative order, for $n\geq 5$, fall into two classes, distinguished by their transformation properties under the compact R-symmetry group $H=SO(3)\times SO(2)$. The first class of supervertices are given by the $SO(3)$ orbits of $\delta^{16}(Q)$, which transforms in the spin $2(n-4)$ representation of the $SO(3)$, and are invariant with respect to the $SO(2)$. The second class of supervertices are the $V_n^\pm$ in (\ref{hspe}). They are charged under the $SO(2)$ and are singlets with respect to the $SO(3)$.

The coefficient of $R^4$ coupling, as a function of the scalar vevs, can be denoted $f(\Omega, U)$. Here $\Omega$ is a real symmetric $3\times 3$ matrix of determinant 1 that parameterizes the coset $SL(3)/SO(3)$ (similarly to the matrix $g_{ij}$ in section 5.1). $U$ is a complex parameter that parameterizes $SL(2)/SO(2)$, which can be identified with the Poincar\'e upper half plane. We will write $U=U_1+iU_2$. 

There are no supervertices at 8-derivative order that transform nontrivially under both the $SO(3)$ and the $SO(2)$ factors of the R-symmetry group. Consequently, 6-point couplings of the form $\delta\Omega \delta U R^4$ do not admit local supervertex completions. They cannot be components of a nonlocal superamplitude either, because such a superamplitude cannot factorize into lower point supervertices, due to the mismatch of R-symmetry representation. We conclude that the coupling $\delta \Omega \delta U R^4$ cannot exist at all, which means that the function $f(\Omega, U)$ splits into a sum of two terms,
\ie
f(\Omega, U) = f_{SL(3)}(\Omega) + f_{SL(2)}(U).
\fe
By the same argument as in type IIB supergravity \cite{Wang:2015jna}, $f_{SL(2)}(U)$ should obey a differential equation of the form
\ie
4U_2^2 \partial_U \partial_{\overline{U}}f_{SL(2)}(U) = a f_{SL(2)}(U),
\fe
for some constant $a$. The second order derivative of $f_{SL(3)}(\Omega)$, on the other hand, transforms as
\ie
{\rm Sym}^2 {\bf 5} = {\bf 1}\oplus {\bf 5}\oplus {\bf 9}
\fe
of the $SO(3)$ R-symmetry.
The 8-derivative 6-point supervertex that contains $(\delta\Omega)^2 R^4$ coupling, as described in section 3.2,  transforms in the 9-dimensional (spin-4) representation of $SO(3)$. By the same factorization argument as in the 7D case, we end up with two sets of differential equations on $f_{SL(3)}(\Omega)$, in the representation ${\bf 1}$ and ${\bf 5}$ respectively. That is, if we expand
\ie
\delta f_{SL(3)}(\Omega) = \delta \Omega_{ij} f^{ij}(\Omega) + \delta \Omega_{ij}\delta\Omega_{k\ell} f^{ij,k\ell}(\Omega) + {\cal O}((\delta\Omega)^3),
\fe
where $f^{ij}$ and $f^{ij,k\ell}$ are restricted to be traceless with respect to $(ij)$ and $(k\ell)$, then we must have relations of the form
\ie
& \Omega_{ij}\Omega_{k\ell} f^{ik,j\ell}(\Omega) = b f_{SL(3)}(\Omega),
\\
& \Omega_{k\ell} f^{ik,j\ell}(\Omega) - {1\over 3} \Omega^{ij} \Omega_{k\ell} \Omega_{mn} f^{mk,n\ell}(\Omega) = c  \left[ f^{ij}(\Omega) - {1\over 3} \Omega^{ij} \Omega_{mn} f^{mn}(\Omega) \right] .
\fe
Comparison with string tree level contributions indicates that
\ie
a=0,~~b=0,~~c=-{5\over 12}.
\fe

In the 1PI quantum effective action of string theory, one encounters a subtlety. Namely, the 1-loop contribution to $R^4$ coupling in 8D has a non-analytic momentum dependence. This non-analyticity in the effective action would be removed if we introduce an IR cut off, or consider a Wilsonian effective action of the massless fields. A priori, the supersymmetry Ward identities are respected by the Wilsonian effective action, rather than the 1PI effective action. An IR cut off would introduce an anomalous transformation of the effective action under the U-duality symmetry of string theory. In  \cite{Green:2010wi}, the function $f(\Omega, U)$ is defined as the $R^4$ coefficient in the 1PI effective action, which is invariant under the U-duality group $SL(3,\mathbb{Z})\times SL(2,\mathbb{Z})$. The price to pay, for insisting on manifest U-duality invariance of the effective action, is that one must modify the naive supersymmetry Ward identity and allow for an anomalous constant term on the RHS of the Laplace equations for $f_{SL(3)}(\Omega)$ and $f_{SL(2)}(U)$.

\subsection{The 9D case}

As described in section 3.1, the 8-derivative $n$-point supervertices in 9D maximal supergravity are given by $\delta^{16}(Q)$ and its CPT conjugate, that contain couplings of the form $(\delta\tau)^{n-4}R^4$ and $(\delta\overline\tau)^{n-4}R^4$, just like in type IIB supergravity. We explained in section 4.2 that there is a special 5-point supervertex at 8-derivative order that is neutral under the $U(1)$ R-symmetry and contains the coupling $\delta\sigma R^4$. On the other hand, there are no 6-point supervertices that contain either $\delta\tau\delta\overline\tau R^4$, $(\delta\sigma)^2 R^4$, or $\delta\tau\delta\sigma R^4$. The absence of these supervertices imply the following constraining relations on the coefficient $f(\tau,\bar\tau,\sigma)$ of the $R^4$ coupling:
\ie
& 4\tau_2^2\partial_\tau \partial_{\overline\tau} f = a_1 \partial_\sigma f + a_2 f,
\\
& \partial_\sigma^2 f = a_3 \partial_\sigma f + a_4 f,
\\
& \partial_\tau \partial_\sigma f = a_5 \partial_\tau f.
\fe
where $a_1,\cdots,a_5$ are constants that are in principle fixed by supersymmetry Ward identities. It suffices to compare with the structure of string tree level and one-loop contributions to determine
\ie
a_1= -{1\over 2},~~~ a_2 = {3\over 7},~~~ a_3 = {3\over 14}, ~~~ a_4 = {27\over 49},~~~a_5 = -{9\over 14}.
\fe
In particular, this determines that $f(\tau,\bar\tau,\sigma)$ takes the form
\ie\label{fff}
f(\tau,\bar\tau,\sigma) = e^{-{9\over 14}\sigma} F_{3\over 4}(\tau,\bar\tau) + e^{{6\over 7}\sigma} C,
\fe
where $F_{3\over 4}$ is an eigenfunction of the Laplacian on the hyperbolic plane parameterized by $(\tau,\bar\tau)$, with eigenvalues $3/4$, and $C$ is a constant. We emphasize that (\ref{fff}) is entirely a consequence of supersymmetry. Combined with $SL(2,\mathbb{Z})$ symmetry that acts on $\tau$, one can then fix $f(\tau,\bar\tau,\sigma)$ completely, as was explained in \cite{Green:1997as,Green:2010wi}.

\subsection{12 and 14-derivative terms}

There is a unique 4-point supervertex at 12-derivative order, $\delta^{16}(Q)(s^2+t^2+u^2)$, that contains couplings of the schematic form $D^4R^4+\cdots$, whose coefficient we denote by $f_4(\phi^I)$, or $f_4(g)$ in the coset notation introduced in the previous subsection. Likewise, there is a unique 4-point 14-derivative supervertex, $\delta^{16}(Q) (s^3+t^3+u^3)$, that contains couplings of the form $D^6R^4+\cdots$, whose coefficient we denote by $f_6(\phi^I)$, or $f_6(g)$. In this section, we analyze the supersymmetry constraints on $f_4$ and $f_6$.

We begin with the analysis of 12-derivative couplings in $D=5,4,3$ dimensions. In these dimensions, the 6-point ${1\over 4}$ BPS supervertices fit in a single irreducible representation of the R-symmetry group. The absence of the singlet 6-point 12-derivative supervertex implies, via the factorization of the 6-point superamplitude, that $f_4$ is an eigenfunction with respect to the Laplacian on the scalar manifold $G/H$. Unlike the 8-derivative coupling constraints, at 12-derivative order, there are no other second order differential constraints on $f_4$. This is consistent with the proposals of \cite{Green:2010wi} and the classification of supersymmetry invariants in \cite{Bossard:2014aea}.

\begin{figure}[htb]
\centering
\includegraphics[scale=1.9]{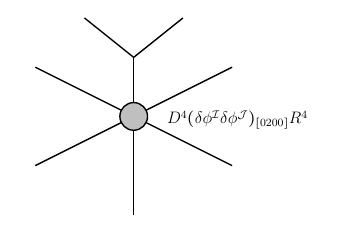}
\caption{Factorization of the five dimensional 7-point 12-derivative superamplitude through one $D^4(\D\phi^{\cal I}\D\phi^{\cal J})_{[0200]}R^4$ supervertex.}
\label{5dD4R4}
\end{figure}

There are, on the other hand, independent {\it third} order differential constraints, that involve $\nabla_{(\cal I}\nabla_{\cal J}\nabla_{\cal K)} f_4$. These follow from the absence of 7-point, 12-derivative supervertices in certain representations of the R-symmetry group (as described in Section 3), and the corresponding factorization of 7-point superamplitudes (see Figure \ref{5dD4R4} for the 5D case). The independently third order differential constraints on $f_4$ are of the form
\ie
& D=5: ~~~ \left.\nabla_{(\cal I}\nabla_{\cal J}\nabla_{\cal K)} f_4\right|_{[0200]} \sim \left.\nabla_{(\cal I}\nabla_{\cal J)} f_4 \right|_{[0200]},~~\left.\nabla_{(\cal I}\nabla_{\cal J}\nabla_{\cal K)} f_4\right|_{[2001]}=0,
\\
& D=4: ~~~\left.\nabla_{(\cal I}\nabla_{\cal J}\nabla_{\cal K)} f_4\right|_{[0200000]}=\left.\nabla_{(\cal I}\nabla_{\cal J}\nabla_{\cal K)} f_4\right|_{[0000020]}=\left.\nabla_{(\cal I}\nabla_{\cal J}\nabla_{\cal K)} f_4\right|_{[1001001]}=0,
\\
& D=3: ~~~\left.\nabla_{(\cal I}\nabla_{\cal J}\nabla_{\cal K)} f_4\right|_{[01000001]}=0.
\fe

Similar higher order differential constraints holds for the coefficient of $D^4R^4$ in higher dimensions. To derive such constraints, we again need to a classification of higher point supervertices. In higher than five dimensions, due to the smaller R-symmetry groups, there appears to be exceptional 1/4 BPS supervertices that do not lie in the R-symmetry orbit of, for instance, (3.17) in eight dimensions. Nonetheless, all 6-point (and higher) 1/4 BPS supervertices can be uplifted from 5D using the prescription of section 4.

\begin{figure}[htb]
\centering
\begin{minipage}{0.3\textwidth}
\centering
\includegraphics[scale=1.2]{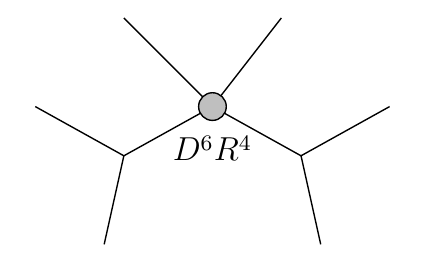}\\
\end{minipage}  
\begin{minipage}{0.3\textwidth}
\centering
\includegraphics[scale=1.2]{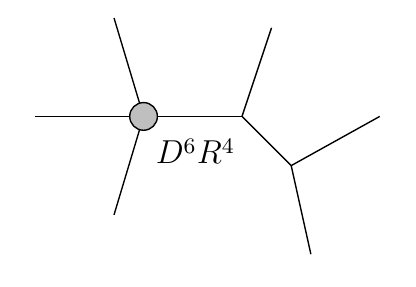}\\
\end{minipage}  
\begin{minipage}{0.3\textwidth}
\centering
\includegraphics[scale=1.2]{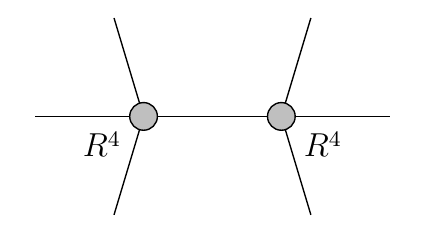}\\
\end{minipage}  
\caption{Factorizations of the 6-point 14-derivative superamplitude through one $D^6R^4$ supervertex or two $ R^4$ supervertices.}
\label{7dD6R4}
\end{figure}

At the 14-derivative order, the absence of the 6-point supervertex in the singlet of R-symmetry implies that the Laplacian $\Delta f_6$ is dictated by the factorization of the R-singlet 6-point superamplitude at the same momentum order. The latter admits two different factorization channels: through the 14-derivative $D^6R^4$ vertex, or through a pair of the 8-derivative $R^4$ vertices as in Figure \ref{7dD6R4}. Consequently, $f_6$ is subject to a second order differential equation of the form
\ie
& \Delta f_6 = a f_6 + b f^2,
\fe
where $f$ is the coefficient of $R^4$ coupling, exactly as in the case of type IIB supergravity \cite{Wang:2015jna} (though the coefficients $a,b$ may differ).

These are not the complete set of supersymmetry constraints on $f_6$, however. The non-singlet components of the 6-point vertex $\delta\phi^{\cal I}\delta\phi^{\cal J} D^6R^4$ do admit local supervertex completions. We expect independent higher order differential constraints, due to the absence of certain 7 and higher point supervertices.\footnote{These are analyzed in \cite{Bossard:2015uga} from the superspace approach.} A full classification of the higher point 14-derivative supervertices will not be attempted here.

\subsection{Higher point supervertices}

The coefficients of $n$-point supervertices for $n>4$ are related to that of the 4-point supervertex at the same derivative order by the soft relations discussed in section 2. For instance, once we determine the 4-point 8-derivative supervertex $f(\phi)\delta^{16}(Q)$, the 5-point supervertex at the same derivative order is given by
\ie\label{phir}
{\partial\over\partial \phi^I} f(\phi)\, \widehat e^I\cdot \left[ \widehat v\, \delta^{16}\Big(\sum_{i=1}^5 Q_i\Big)\right]^{SO(5)_R}
\fe
Here $\widehat e^I$ and $\widehat v$ are auxiliary tensors in the $[2,0]$ representation space of $SO(5)_R$. The $\widehat e^I$ is a set of unit basis tensors, and $\widehat v$ is the highest weight state of $SO(5)_R$ in the representation $[2,0]$, while the 5-point supervertex $\delta^{16}(Q)$ is the lowest weight state in its $SO(5)_R$ orbit. The superscript in (\ref{phir}) stands for the average over the $SO(5)_R$ rotation on $\widehat v$ and $\delta^{16}(Q)$ simultaneously. In the vacuum where $\phi$ acquires expectation value $\phi_0$, the supervertex (\ref{phir}) contains scalar-graviton couplings of the form $\partial_{\phi^I} f(\phi_0)\, \delta\phi^I R^4$.


The soft relations fix all the 8-derivative supervertices in terms of the 4-point supervertex $f(\phi)\delta^{16}(Q)$. Likewise, all the 12-derivative supervertices in terms of the 4-point supervertex $f_4(\phi)\delta^{16}(Q)(s^2+t^2+u^2)$. These 12-derivative supervertices are linear combinations of $SO(5)$ rotations of $\delta^{16}(Q)\sum_{i<j} s_{ij}^2$.

At 14-derivative order, there are a set of new independent $n$-point supervertices for $n\geq 6$. This is because $\delta^{16}(Q)\sum_{i<j} s_{ij}^3$ and $\delta^{16}(Q)\sum_{i<j<k} s_{ijk}^3$, where $s_{ijk}\equiv -(p_i+p_j+p_k)^2$, are generally independent for $n\geq 6$\cite{Elvang:2010jv}. While the sum of the coefficients of these two 6-point couplings is determined by the soft limit, in terms of derivatives of the coefficient $f_6(\phi)$ of $D^6R^4$, the individual 6-point coefficients are not fixed by such a relation. The argument based on factorization of 8-point superamplitudes at 14-derivative order indicates that the coefficients of $\delta^{16}(Q)\sum_{i<j} s_{ij}^3$ and $\delta^{16}(Q)\sum_{i<j<k} s_{ijk}^3$ should still obey second order differential equations in $\phi^I$, whose sources are quadratic in derivatives of $f(\phi)$.

Let us illustrate the supersymmetry constraints on these 6-point couplings, in the example of a type IIB supergravity theory in ten dimensions. Suppose we have the following 6-point supervertex at 14-derivative order,
\ie
\delta^{16}(Q) \left[ F_1(\tau,\bar\tau) \sum_{1\leq i<j\leq 6} s_{ij}^3 + F_2(\tau,\bar\tau) \sum_{1\leq i<j<k\leq 6} s_{ijk}^3 \right],
\fe
corresponding to couplings of the schematic form $D^6 (\delta\tau)^2 R^4+\cdots$.
Taking the soft limit on $p_6$, we obtain the 5-point supervertex
\ie
\delta^{16}(Q) \left[ F_1(\tau,\bar\tau) + 2F_2(\tau,\bar\tau) \right] \sum_{1\leq i<j\leq 5} s_{ij}^3.
\fe
Thus, from the soft relation discussed in section 2, we learn that
\ie
F_1(\tau,\bar\tau) + 2F_2(\tau,\bar\tau) = \nabla_\tau^2 f_6(\tau,\bar\tau) = (\partial_\tau^2 - {i\over \tau_2}\partial_\tau) f_6(\tau,\bar\tau),
\fe
where $f_6$ is the coefficient of the 4-point supervertex $\delta^{16}(Q)\sum_{1\leq i<j\leq 4} s_{ij}^3$, corresponding to the coupling $D^6R^4+\cdots$.

\begin{figure}[htb]
\centering
\begin{minipage}{0.45\textwidth}
\centering
\includegraphics[scale=1.4]{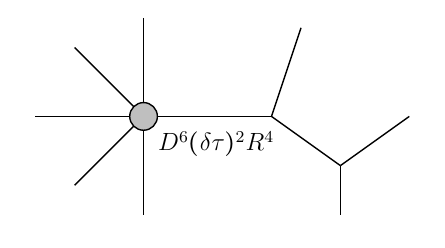}\\
\end{minipage}  
\begin{minipage}{0.45\textwidth}
\centering
\includegraphics[scale=1.4]{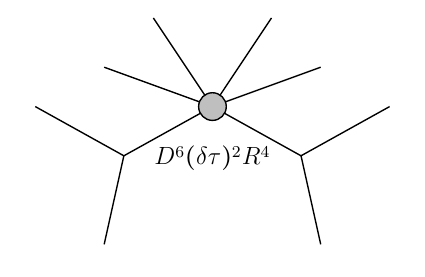}\\
\end{minipage}  
\begin{minipage}{0.45\textwidth}
\centering
\includegraphics[scale=1.4]{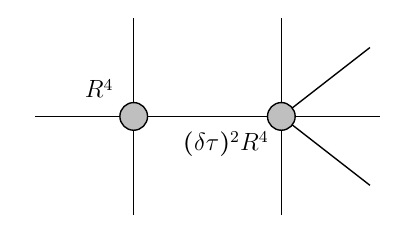}\\
\end{minipage}  
\begin{minipage}{0.45\textwidth}
\centering
\includegraphics[scale=1.4]{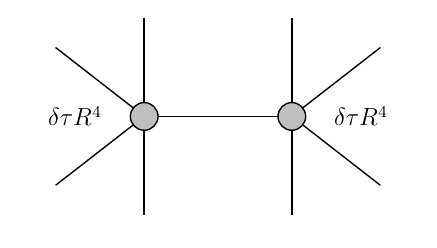}\\
\end{minipage}  
\caption{Factorizations of the 8-point 14-derivative superamplitude through $D^6(\D\tau)^2 R^4$ supervertices, one $R^4$ supervertex and one $(\D\tau)^2 R^4$ supervertex, and a pair of $\D\tau R^4$ supervertices respectively.}
\label{7dphi2D6R4}
\end{figure}

In order to further constrain $F_1$ (or $F_2$), we need to consider the 8-point amplitude with an extra pair of $\tau, \overline\tau$ emissions. The absence of a local supervertex of this form leads to the relation
\ie\label{sixf}
\Delta F_1 = a F_1 + b \nabla_\tau^2 f_6 + c f_0 \nabla_\tau^2 f_0 + d (\partial_\tau f_0)^2,
\fe
where the RHS come from the possible factorization channels of the 8-point superamplitude (Figure \ref{7dphi2D6R4}). $f_0(\tau,\bar\tau)$ is the coefficient of the $R^4$ supervertex. $\Delta$ is the Laplacian operator on a covariant tensor of weight $(2,0)$ in $(\tau, \bar\tau)$, on the hyperbolic plane. A priori, $a,b,c,d$ are constants fixed by supersymmetry Ward identities. One may try to determine these constants by comparison with string perturbation theory. However, unlike the differential equation for $f_6$, which involves two coefficients that can be fixed by comparison with tree level and one-loop string amplitudes, to fully determine the coefficients in (\ref{sixf}) may require knowing the explicit contributions to the 6-point amplitude at 14-derivative order from up to 3-loop string amplitudes. Alternatively, one may try to solve the supersymmetry Ward identities on a general 6-point nonlocal superamplitude at 14-derivative order directly. We leave this to future work.

\section{Summary and discussions}

In this paper we gave a conjectural classification of ${1\over 2}$ BPS supervertices in maximal supergravity theories in spacetime dimension $D$ for $3\leq D\leq 9$. In dimension 7 and below, all $n$-point 8-derivative supervertices are in the R-symmetry orbit of $\delta^{16}(Q)$. In $D=8$, there is an additional set of $n$-point supervertices $V_n^\pm$, that transforms under the $SO(3)\times SO(2)$ R-symmetry group as singlets of $SO(3)$ and charged under the $SO(2)$. In $D=9$, there is an exceptional 5-point supervertex that is neutral under the $U(1)$ R-symmetry.

Families of ${1\over 4}$ BPS supervertices are constructed as well, starting at 6-point, 12-derivative order. Our construction appears to be exhaustive in spacetime dimensions 5 and below. We also gave examples of ${1\over 8}$ BPS supervertices, although they have not been classified in detail.

We further showed that among the ${1\over 2}$ BPS supervertices of 9D supergravity, only the 4-point supervertices of the form $\delta^{16}(Q){\cal F}(s,t,u)$ can be lifted to Lorentz invariant supervertices in 11 dimensions. It is likely that they exhaust all F-term supervertices in 11D, although this remains to be proven. If this is true, it would imply that the only F-term supervertices that control the M-theory effective action in eleven dimensions are 4-point supervertices that contain $R^4$, $D^4R^4$, and $D^6R^4$ couplings.\footnote{The uplifting procedure from 9D also assumes that the number of independent momenta in the vertex is no more than 9. So our construction of 11D supervertices by uplifting from 9D a priori only applies to $n$-point supervertices for $n\leq 10$.} The existence of such couplings and their coefficients in the M-theory effective action have been previous established \cite{Green:2005ba}. Our argument for the non-renormalization conditions, based on factorization of superamplitudes, then suggests that all M-theory amplitudes up to 14-th order in the momentum expansion can be fixed by the coefficients of these supervertices. In principle, this should determine up to $R^7$ terms in the derivative expansion of the M-theory effective action. In practice, one may construct the exact 11 dimensional amplitudes at these orders in the momentum expansion by uplifting perturbative string amplitudes in lower dimensions.

The main application of the classification of F-term supervertices in this paper is the derivation of the non-renormalization conditions on the moduli dependence of F-term couplings, such as $f(\phi) R^4$, $f_4(\phi) D^4R^4$, and $f_6(\phi) D^6R^4$, in lower dimensional maximal supergravity theories. We find that for $3\leq D\leq 9$, besides an equation that asserts $f(\phi)$ is an eigenfunction of the Laplacian on the scalar manifold $G/H$, as was proposed in \cite{Obers:1999um,Green:2010wi}, there are additional constraints on the Hessian of $f(\phi)$. We verified explicitly in dimensions 6 and above that these constraining equations on the $R^4$ coupling coefficient are precisely consistent with previous proposals \cite{Obers:1999um,Green:2010wi} in toroidally compactified type II string theory, based on automorphic functions of the U-duality group.  

The constraints on $f_4$ is more intricate. Besides the condition that $f_4$ is an eigenfunction of the Laplacian, there are no other second order differential constraints on $f_4$, as we have seen through the explicit construction of 6-point, 12-derivative supervertices in dimensions $D=5,4,3$. On the other hand, $f_4$ is subject to a set of third order differential equations. Our results are consistent with that of \cite{Bossard:2014aea}, where the same problem is analyzed using harmonic superspace.

We pointed out that at 14-derivative order, there is a new set of F-term supervertices that arise at 6-point order and higher, due to independent supervertices of the form $\delta^{16}(Q)\sum s_{ij}^3$ and $\delta^{16}(Q)\sum s_{ijk}^3$. The structure of supersymmetry constraints on these couplings was discussed in section 5.3, but the precise coefficients in these equations are not yet fixed. Without trying to solve supersymmetry Ward identities on 8-point superamplitudes directly, to fully determine these equations requires more input from string perturbation theory (involving 6-point amplitudes). This is an interesting problem that could provide new tests of string perturbation theory at higher loop order and S-duality, which we leave to future work.

Thus far, we have little to say about the D-terms, whose coefficients appear to be unconstrained as functions of the moduli fields. Let us contrast our result with that of the Abelian effective action on the Coulomb (or tensor) branch of maximally supersymmetric gauge theories (or the 6D $(2,0)$ theory) \cite{Paban:1998ea,Paban:1998qy,Sethi:1999qv,Lin:2015zea,Lin:2015ixa,Cordova:2015vwa}. In the gauge theory context, there are 16 rather than 32 supersymmetries, and F-terms only arise at 4 and 6-derivative orders. However, when the theory at the origin of the Coulomb branch is a superconformal theory, the Coulomb branch effective action is also controlled by spontaneously broken conformal symmetry \cite{Schwimmer:2010za,Komargodski:2011vj,Elvang:2012st}, which introduces strictly stronger constraints than supersymmetry alone. Do the effective actions of maximal supergravity or toroidally compactified type II string or M-theory admit similar hidden symmetries? Proposals on the exact D-term couplings, such as $D^8R^4$, were made in \cite{Green:2005ba,Basu:2006cs,Basu:2008cf,Basu:2013goa,D'Hoker:2015foa}. It is not clear to us why such couplings would be subject to non-renormalization conditions. If they are, then some yet unknown hidden symmetry may be at play.

\bigskip

\section*{Acknowledgments}

We would like to thank Clay C\'ordova, Thomas Dumitrescu, Henriette Elvang, and Shu-Heng Shao for discussions. We are grateful to Guillaume Bossard for pointing out the significance of higher order differential constraints. Y.W. is supported in part by the U.S. Department of Energy under grant Contract Number  DE-SC00012567. X.Y. is supported by a Sloan Fellowship and a Simons Investigator Award from the Simons Foundation.

\bibliography{msugra} 
\bibliographystyle{JHEP}

\end{document}